\documentclass[structabstract]{aa}

\usepackage{graphicx}
\usepackage{array}
\usepackage{lscape}
\usepackage{rotating}
\usepackage{longtable}
\usepackage{multirow}
\usepackage{color}
\usepackage{amsfonts}
\usepackage{amsmath}

\usepackage{natbib}

\def\lt{LiTE}
\def\oc{\textit{O-C}}

\begin{document}

        \title{New analysis of the light time effect in TU Ursae Majoris}

        \author{Li\v{s}ka, J.\inst{1}, Skarka, M.\inst{1,2}, Mikul\'{a}\v{s}ek, Z.\inst{1}, Zejda, M.\inst{1}, \& Chrastina, M.\inst{3}
          }
        \offprints{J.~Li\v ska,\\  \email{jiriliska@post.cz}}

        \institute{Department of Theoretical Physics and Astrophysics, Faculty of Science, Masaryk University, 
                                                Kotl\'a\v rsk\'a 2, CZ-611~37 Brno, Czech Republic, \email{jiriliska@post.cz}    
        \and
        Konkoly Observatory, Research Centre for Astronomy and Earth Sciences, Hungarian Academy of Sciences, H-1121 Budapest, Konkoly Thege Mikl\'{o}s \'{u}t 15-17, Hungary,
        \and    
        independent scientist, http://chrastina.kozmos.sk, Brusla\v{r}sk\'{a} 959/6, CZ-102 00 Praha, Czech Republic
        }   
   \date{Received ...;  accepted }

 \abstract
{Recent statistical studies prove that the percentage of RR Lyrae pulsators that are located in binaries or multiple stellar systems is considerably lower than might be expected. This can be better understood from an in-depth analysis of individual candidates. We investigate in detail the light time effect of the most probable binary candidate \object{TU UMa}. This
is complicated because the pulsation period shows secular variation.}
{We model possible light time effect of \object{TU UMa} using a new code applied on previously available and newly determined maxima timings to confirm binarity and refine parameters of the orbit of the
RRab component in the binary system. The binary hypothesis is also tested using radial velocity measurements.} 
{We used new approach to determine brightness maxima timings based on template fitting. This can also be used on sparse or scattered data. This approach was successfully applied on measurements from different sources. To determine the orbital parameters of the double star \object{TU UMa}, we developed a new code to analyse light time effect that also includes secular variation in the pulsation period. Its usability was successfully tested on \object{CL Aur}, an eclipsing binary with mass-transfer in a triple system that shows similar changes in the \oc~diagram. Since orbital motion would cause systematic shifts in mean radial velocities (dominated by pulsations), we computed and compared our model with centre-of-mass velocities. They were determined using high-quality templates of radial velocity curves of RRab stars.}
{Maxima timings adopted from the GEOS database (168) together with those newly determined from sky surveys and new measurements (85) were used to construct an \oc~diagram spanning almost five proposed orbital cycles. This data set is three times larger than data sets used by previous authors. Modelling of the \oc~dependence resulted in 23.3-year orbital period, which translates into a minimum mass of the second component of about 0.33\,$\mathfrak{M}_{\odot}$. Secular changes in the pulsation period of \object{TU~UMa} over the whole \oc~diagram were satisfactorily approximated by a parabolic trend with a rate of $-2.2$\,ms\,yr$^{-1}$. To confirm binarity, we used radial velocity measurements from nine independent sources. Although our results are convincing, additional long-term monitoring is necessary to unambiguously confirm the binarity of \object{TU UMa}.}
{}

\keywords{stars: variables: RR Lyrae -- binaries: general -- methods: data analysis -- techniques: photometric -- techniques: radial velocities -- stars: individual: TU UMa} 
         \authorrunning{Li\v{s}ka et al.}
         \titlerunning{Light time effect in TU Ursae Majoris}
        
   \maketitle
%
%

\section{Introduction}\label{introductionsec}

A significant part of stars are located in double or multiple stellar systems. However, reviews of pulsating stars bound in binaries \citep[e.g.][]{szatmary1990,zhou2010} clearly show the lack of stellar pairs with an RR Lyrae component. The number of currently confirmed binaries comprising an RR Lyrae type pulsator can be counted on one hand.

The binarity of an object can be revealed in many different ways. For example, detection of eclipses, periodic radial velocity (RV) changes, or regular astrometric shifts in a visual binary can serve as a direct proof of binarity. A companion of a periodic variable star can also be detected indirectly through changes in timings of light extrema, the so-called light time effect (hereafter \lt). RR Lyrae stars are generally located at larger distance from Earth, hence astrometric detection of binarity is highly unlikely. Since the spectra of RR Lyrae stars are influenced by pulsations, discovering the binary nature of stars through changes in the position of spectral lines is also difficult \citep[e.g.][]{fernley1997,solano1997}. Thus the most promising methods are detection of eclipses and \lt. 

In the Large Magellanic Cloud (LMC) three candidates for RR Lyraes in eclipsing binaries were detected \citep{soszynski2003}. However, these objects were identified as optical blends consisting of two objects, RR Lyrae star and eclipsing system \citep{soszynski2003,prsa2008}. A very interesting object was identified by \citet{soszynski2011} and subsequently studied by \citet{pietrzynski2012} and \citet{smolec2013}. This peculiar eclipsing system with an orbital period of 15.24\,d contains a component that mimics an RR Lyrae pulsator. The detailed study showed that this object, \object{OGLE-BLG-RRLYR-02792}, has a very low mass of only 0.26\,$\mathfrak{M}_{\odot,}$ which is too low for classical RR Lyrae stars. Other physical characteristics also indicate that this binary component is a very special object as a result of an evolution in a close binary followed by a life similar to a classical RR Lyrae star. \citet{pietrzynski2012} included the object in a new class of pulsating stars called binary evolution pulsators (BEP). \object{OGLE-LMC-RRLYR-03541} is another candidate for RR Lyr in eclipsing binary \citep{soszynski2009}. Nonetheless, the similarity in the shape of the light curve and the short orbital period of 16.229\,d might mean that it belongs to the class of BEP \citep{hajdu2015b}. In addition, it is not yet excluded that it could be actually a blend of two close stars.

\citet{szatmary1990} published a list of various types of pulsating stars bound in binary systems that were visually identified on the basis of the \lt~in their \oc~diagrams. However, detailed information and references are missing. The catalogue of various binary systems with pulsating components from \citet[][version December 2014]{zhou2010} contains in the RR Lyrae class \object{TU UMa}, \object{OGLE-BLG-RRLYR-02792}, and several tens of candidates without any closer information. Therefore, including some of these objects as binary systems (in both catalogues) is at least doubtful.

Several examples of RR Lyrae stars in binaries were very recently
identified through the analysis of their \oc~diagrams. \citet{li2014} found that \object{FN~Lyr} and \object{V894~Cyg} are probably in pairs with brown dwarves. \citet{hajdu2015} found 20 additional candidates among OGLE bulge RRab variables. \citet{liska2015} presented an analysis of 11 new binary candidates located in the Galactic field. Nevertheless, all these candidates need to be confirmed spectroscopically or using another independent method.

In this paper, we present a new analysis of the \lt~in \object{TU~UMa} \citep[23-year long cycle, detected by][]{szeidl1986} that is based on a much wider sample of \oc~than was used for the \lt~in \object{TU~UMa}  before \citep{wade1999}. Highly accurate photometric observations that cover two thirds of the proposed orbital period, are newly available. In Sect.~\ref{historysec} we briefly discuss the history of observation of \object{TU~UMa} with an emphasis on its binary nature. In Sect.~\ref{datasec} we summarise the characteristics of our data sample. Except for data from various sources, we used the original measurements gathered in 2013--2014. We apply our \lt~procedure (described in Sect.~\ref{modelsec} and Appendix~\ref{liteappendix1}) and model the \object{TU~UMa} data in Sect.~\ref{resultstuumasubsec}. Other proofs for binarity are discussed in Sect.~\ref{proofsec}, and all results are summarised in Sect.~\ref{summarysec}.

\section{History of \object{TU UMa} observations}\label{historysec}

TU Ursae Majoris = \object{AN 1.1929} = \object{BD+30 2162} = \object{HIP 56088} ($\alpha$ = 11$^{\rm h}$29$^{\rm m}$48\fs49, $\delta$ = +30\degr04\arcmin02\farcs4, J2000.0) is a pulsating RR Lyrae star of Bailey ab type. According to the Variable Star Index\footnote{http://www.aavso.org/vsx/} \citep{watson2006}, its brightness in $V$ band varies in the range of 9.26--10.24\,mag (spectral type A8--F8) with a period of about 0.558\,d. No signs of the Bla\v{z}ko effect have been reported for \object{TU UMa} up to now. 

The variability of \object{TU UMa} was discovered by \citet{guthnick1929} on Babelsberg plates. Many authors thereafter studied the star using photoelectric photometry and spectroscopy. A detailed description of the history of \object{TU UMa} research was performed by \citet{szeidl1986}. Only the most important information about the \lt~is briefly mentioned below, since about 180 articles with the keyword \object{TU~UMa} are currently retrievable at the NASA ADS portal. 

\citet{payne-gaposchkin1939} was the first who noted cyclic variations in maxima timings and proposed a 12400-day (34-year) long cycle. Important results were obtained by \citet{szeidl1986}, who mentioned a secular period decrease that causes the parabolic trend in
the \oc~diagram, and a probable 23-year (8400\,d) variation that
is possibly caused by the binarity of the star. \citet{saha1990a} detected systematic shifts in RVs that indicate binarity, but they used very few RV measurements. They modelled the \lt~for the first time and determined orbital period of 7374.5\,d (20.19\,yr). Their analysis showed that the proposed stellar pair has an extremely eccentric orbit with $e\!=\!0.970$. They considered only a constant pulsation period in their model. The effect of neglecting the secular changes of $e$, $a\,\sin i$, and $\mathfrak{M}_{2}\,\sin^{3} i$ was tested by \citet{wade1992}.

The \lt~with secular variation in the pulsation period was firstly solved by \citet{kiss1995}, who determined more accurate orbital elements and derived an orbital period of 8800\,d (24.1\,yr). \citet{wade1999} collected all available maxima timings and also RVs and successfully verified results from \citet{saha1990a}. They obtained five different groups of models of the \lt~with respect to different subsets of maxima timings (all maxima without visual values, only photoelectric and CCD values, etc.). They derived orbital periods in the range from 20.26\,yr to 24.13\,yr depending on the particular data set and the number of fitting parameters (with or without parabolic trend). Subsequently, they used nine derived sets of orbital elements to reconstruct the orbital RV curve of the pulsating component and to compare it with shifts in observed RVs. Since then, \object{TU UMa} has not been
studied with regard to its LiTE for about 15 years. However, improvements of quadratic ephemeris were performed by \citet{arellanoferro2013},
for instance. Many high-accurate maxima timings (mainly CCD measurements) were published during this 15-year interval. The currently available data exceed five proposed orbital cycles.

\section{Data sources}\label{datasec}

\subsection{GEOS RR Lyrae database}\label{geossubsec}
Since the GEOS RR Lyrae database\footnote{http://rr-lyr.irap.omp.eu/dbrr/} \citep[{\it{Groupe Europ\'{e}en d\textquoteright Observations Stellaires}}, ][]{boninsegna2002,leborgne2007} is the most extended archive of times of RR Lyrae star maxima, we used this as the main data source for our analysis. The version of the database of November 2014 contains corrected values for three maxima timings from \citet{kiss1995} that originally contained incorrect heliocentric corrections \citep{wade1999}. 

We discarded all visually determined maxima timings because they
were inaccurate and omitted two photographic maxima from \citet{robinson1940} and \citet{payne-gaposchkin1954}, which were replaced by our maxima from the DASCH project (see Sect.~\ref{othersourcessubsec}). We used also the latest maxima timings from \citet{hubscher2014,hubscher2015} that were not included in the version of the GEOS list we used.

We paid special attention to maxima timings based on data from sky surveys such as Hipparcos \citep{esa1997} or ROTSE = NSVS \citep{wozniak2004} for the GEOS values\footnote{GEOS data marked \textquoteleft
pr. com.\textquoteright~were not used.}. These timings were acquired
with a special method because in most cases they were determined statistically based on the combination of many points that are often spread out over a few years. \oc~values determined from such data sets very often did not follow the general trend of \oc~dependence\footnote{the original value of the maximum timing 2448500.0710 HJD from the Hipparcos satellite \citep{maintz2005},
e.g., has a residual value \oc$_{\rm res} = 0.0096$\,d based on model 2 in Sect.~\ref{resultstuumasubsec}, but the standard deviation of CCD measurements determined from the model is only 0.0017\,d.}. They were therefore omitted. However, we re-analysed the original data of these surveys (Sect.~\ref{othersourcessubsec}). All used maxima timings are available at the CDS portal.

\subsection{Our observations}\label{observsubsec}

\begin{figure}[t]
\centering
\includegraphics[width=1.00\hsize]{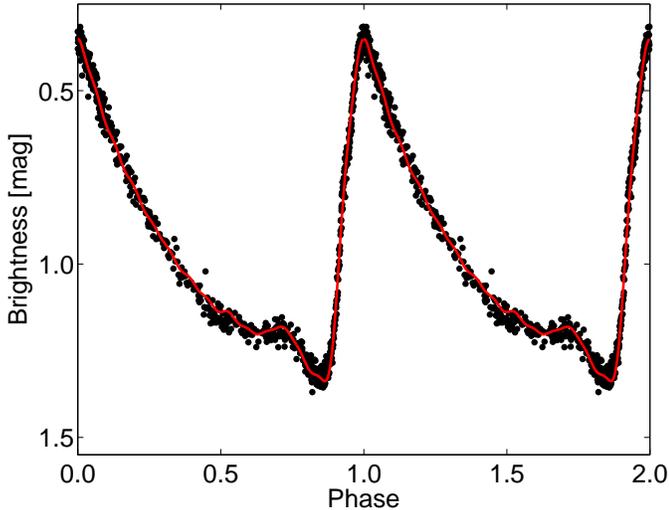}
\caption{Differential magnitudes of \object{TU UMa} obtained with the 1-inch telescope folded with the pulsation period (black dots) plotted together with the model of $V$-band observations from \citep{liakos2011a}.}
\label{Fig:LightCurve}
\end{figure}

As an extension of the GEOS data we also used ten new maxima timings gathered by J.~Li\v{s}ka in 19 nights between December 2013 and June 2014. CCD photometric measurements were performed using two telescopes -- three nights with a 24-inch Newtonian telescope ($vby$ Str\"{o}mgren filters) at Masaryk University Observatory in Brno, and 16 nights with a small 1-inch refractor \citep[{\it green} filter with similar throughput as the Johnson $V$ filter,][]{liska2014a} at a private observatory in Brno. For the small-aperture telescope, five frames with exposure times of 30\,s each were combined to a single image to achieve a better signal-to-noise ratio. The time resolution of such a combined frame is about 170\,s. The comparison star \object{BD+30~2165} was the same for both instruments, but the control stars were \object{BD+30~2164} (for the 24-inch telescope) and \object{HD~99593} (for the 1-inch telescope). Maxima timings were determined via polynomial fitting\footnote{Used for observations with 24-inch telescope where a full light curve was not available.} or the template fitting described in Sect.~\ref{othersourcessubsec}. Monitoring with the small telescope resulted in the well-covered phase light curve shown in Fig.~\ref{Fig:LightCurve}. Except for maxima timings determination, the observations were also
important for detecting possible eclipse (Sect.~\ref{eclipsesec}). Our measurements are available at the CDS portal.

\subsection{Other sources}\label{othersourcessubsec}

We used high-cadence measurements from the SuperWASP project \citep{pollacco2006,butters2010} and Pi of the Sky project \citep[e.g.][]{burd2004,siudek2011} to determine maximum timings from the individual well-covered nights. 

In addition, to extend the \oc~dataset as far as possible, we also analysed data from other large sky surveys (Hipparcos, NSVS) and from the project DASCH \citep[photometry from scanned Harvard plates, e.g.][]{grindlay2009}. For these very sparse but very extended data, maxima timings were estimated using template fitting. The same process was applied to the data from the other sources. 

Firstly, we chose the dataset with the best-quality data and with the best phase-coverage. Since the data from all surveys were of insufficient quality to construct the template light curve (e.g. in Hipparcos and NSVS data the phase around maximum brightness was not observed), $V$-band measurements from \citet{liakos2011a} were used. We then modelled the shape of the light curve $m(t)$ in Matlab with a non-linear least-squares method (hereafter LSM) with an $n$-order harmonic polynomial
\begin{equation}\label{eqfouriersum}
m(t) = A_{0} + \sum_{j=1}^{n} A_{j} \cos\left( 2\,\pi\,j\,\frac{t - M_{0}}{P_{\rm puls}} + \phi_{j}\ \right),
\end{equation}
where $A_{0}$ is the zeroth level of brightness in mag, $A_{j}$ are amplitudes of the components in mag, $t$ is the time of observation in Heliocentric Julian Date (HJD), $M_{0}$ is the zero epoch of maximal brightness in HJD, $P_{\rm puls}$ is the pulsation period in days, and $\phi_{j}$ represents the phase shifts in radians. After some experiments, we found that a polynomial with $n=15$ is sufficient for a good template model. 

Nevertheless, using one template curve for various surveys brings some particularities. Firstly, each dataset has its own magnitude zero level, and the model has to be scaled because different surveys use different filters. Therefore we used the whole light curves for our fit. Individual amplitudes $A_{j}$ and phase shifts $\phi_{j}$ known from the template curve remained fixed, but the total amplitude was rescaled using the ratio of total amplitudes for the analysed and template datasets. The factor was typically
close to 1. In addition, pulsation period and zero epoch had to be slightly refined for each dataset. Subsequently, outliers were iteratively removed.

Individual observational instruments have different spectral sensitivities, therefore it was necessary to verify the usability of the same $V$-band template for different datasets. We used photometric measurements in \textit{UBVRI} filters from \citet{liakos2011a} and compared all curves. The main difference is in amplitude, which is highly dependent on the mean wavelength (the amplitude decreases from 1.3\,mag in $U$ to 0.6\,mag in $I$). When the amplitudes are normalised, the differences between shapes of $U$, $B$, $V$, and $R$ curves are almost negligible (comparable with the noise level). The same applies for the times of maxima. Only $I$-band light changes differ, apparently. Surveys typically observed in broad band filters and mean wavelengths are not known, or the values were only roughly estimated. We can fortunately
estimate them using comparison between amplitudes of survey data and \textit{UBVRI}-amplitudes. We found that effective wavelengths for all surveys lie between the wavelengths of the $B$ and $R$ filters; this means that our approach ($V$-band template) is applicable.

\begin{figure}[t]
\centering
\includegraphics[width=1.00\hsize]{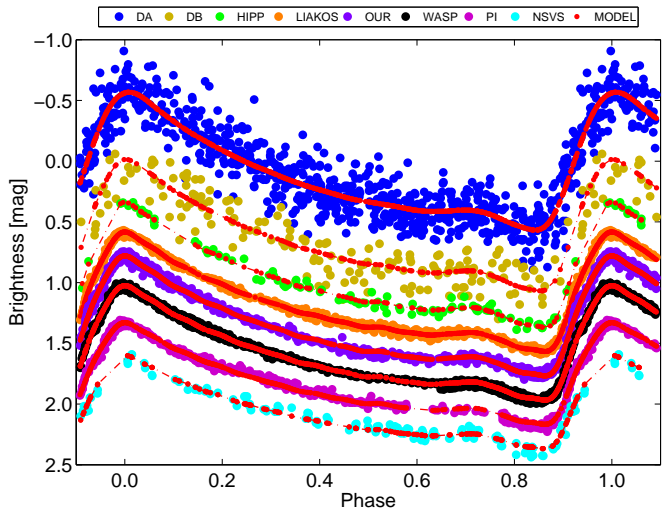}
\caption{Light curves of \object{TU UMa} from different datasets: DA -- DASCH A, DB -- DASCH B, HIPP -- Hipparcos, LIAKOS -- \citet[][$V$-filter]{liakos2011a}, OUR -- this paper (1-inch, $green$-filter), WASP -- SuperWASP (CCD 144), PI -- Pi of the Sky, NSVS -- NSVS, MODEL -- used template curves (DASCH templates are described in Sect.~\ref{DASCHsubsec}). Light curves are vertically shifted to better display the light variability.}
\label{Fig:LCSurveys}
\end{figure}

The good consistence of our template and the observed curves was controlled visually (see Fig.~\ref{Fig:LCSurveys}). For SuperWASP data, which were obtained using six different CCD cameras, a very careful selection of the best nights compared to the template was performed to avoid significant trends that are present in the data.

When the data were sparse without well-defined maxima, it was necessary to divide the whole dataset into smaller subsamples containing typically about 30 points, with a time span from several days to hundreds of days. The data in these subsamples were then compared with the refined template light curve, and the time of maximum was determined. With this subsampling we were of course
only able to estimate the mean time of maximum for the time interval of particular subset. However, in comparison with the total time span of the \oc~values, which cover several decades, this method is fully appropriate. The numbers of all used maxima timings from particular surveys are listed in Table~\ref{NumberMaximaTable}.

\begin{center}
\begin{table}[t]
\centering
\caption{Numbers of new maxima timings of \object{TU UMa} determined from individual projects and from our observations.}\label{NumberMaximaTable}
{\tiny
\def\arraystretch{1.5}
\tabcolsep=3.0pt
\begin{tabular}{ccccccc}
\hline\hline
Hipparcos       & NSVS  & Pi of the Sky & DASCH & SuperWASP     & Our\\
\hline
3                       & 4             & 5                             & 29      & 21            & 10 \\
\hline
\end{tabular}}  
\end{table}
\end{center}

In addition, we determined five maxima times from the data that were omitted or only poorly determined from the analysis in \citet{boenigk1958}, \citet{liakos2011a,liakos2011b}, \citet{liu1989}, and \citet{preston1961}.

The uncertainty of each time of maximum determined by the template fitting method is influenced mainly by the choice of the used model (selection of polynomial degree), the time determined for
the template maximum, and by the quality of the fitted datasets. The first two components were estimated statistically from the template light curve, and their combination was set to 0.0008\,d. The influence of the third source of uncertainties was individually
estimated for each light curve directly from the LSM.

\subsection{Photographic measurements -- DASCH maxima timings}\label{DASCHsubsec}
\begin{figure}[t]
\centering
\includegraphics[width=1.00\hsize]{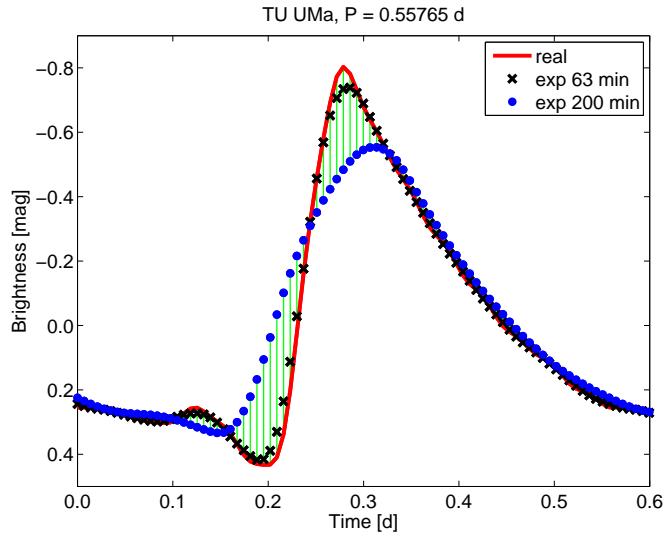}
\caption{Change of the light curve shape caused by long exposure time. The model of the real light curve based on $B$-band observations from \citet{liakos2011a} (red line) is plotted together with models of observed curves with exposures of 63\,min (black crosses) and 200\,min (blue dots).}
\label{Fig:LightCurveExp}
\end{figure}

Photographic measurements that were obtained with exposure times of an hour and longer are special cases. 
Light curves resulting from these observations differ from real curves in amplitude and also in time of minima and maxima (for example the models of $B$-band measurements of \object{TU UMa} from \citet{liakos2011a} in Fig.~\ref{Fig:LightCurveExp}). We performed a detailed analysis of this problem and will publish the results elsewhere. Therefore we discuss only the most important findings. The extrema timings and amplitude change almost linearly with the duration of the exposure time. Maxima are delayed, minima occurred sooner, and
the amplitude is lower than for shorter exposures.

\begin{figure}[t]
\centering
\includegraphics[width=1.00\hsize]{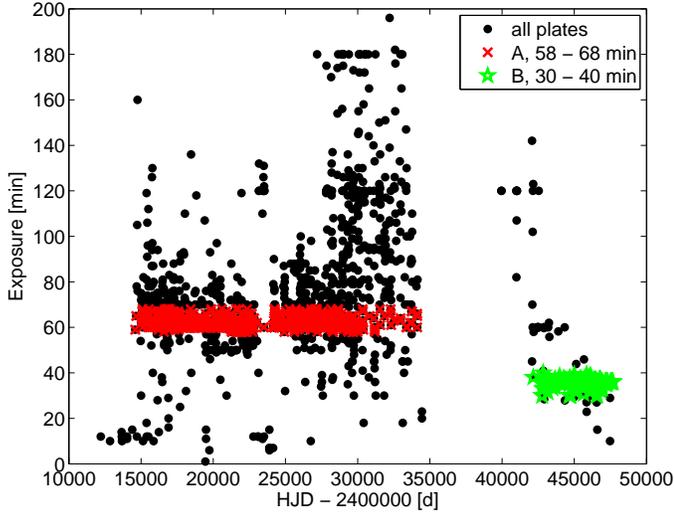}
\caption{Exposure times of the photographic plates of \object{TU UMa} in the DASCH project. The data were divided into two groups according to their integration times -- group A with 58--68\,min (red crosses) and B with 30--40\,min (green stars).}
\label{Fig:DaschExp}
\end{figure}

The photographic plates that were digitalised in the DASCH project
have various integration times, mostly from 1 to 200\,min (Fig.~\ref{Fig:DaschExp}), therefore it is very difficult to correctly analyse the data (different models of the observing curve have to be used). We solved this problem by selecting the measurements with similar exposures (they can be approximated with the same template curve; the time difference is no more than 0.0005\,d).

At first, we selected the measurements according to the exposure -- group A (58--68\,min, older measurements) and group B (30--40\,min, newer measurements). Then we calculated template curves for the mean exposures 63\,min and 35\,min (deformed by the exposure) based on measurements of \citet{liakos2011a}. These two templates were used for the DASCH A and B dataset.

Our tests showed that for $\sim$\,60\,min exposure, for instance, the maximum brightness based on exposure-deformed light curve is delayed of about +0.004\,d in comparison with real changes. This difference will be apparent, for example, in maxima timings determined using polynomial fitting. Therefore, this shift has to be included in the final maxima times. When the template fitting method is applied with an improved time-corrected model (template from the deformed curve), the exposure-length discrepancy is fully reduced to zero.

\section{Modelling the LiTE}\label{modelsec}

\subsection{Light time effect}\label{modellitesubsec}

The \lt~was suggested at the end of the ninteenth century in
the \object{Algol} system by \citet{chandler1888}, while the first detailed theoretical analysis of the problem was performed by \citet{woltjer1922}. \citet{irwin1952a} solved important equations for a part of the direct solution of the \lt~and described a graphical way to determine
the orbital elements. 

Currently, the \lt~is usually solved very accurately by applying equations of motion in a two-body system (the equations from \citet{irwin1952a} included a direct solution of Kepler's equation) or using numerical calculations of a perturbed orbit in a multiple system. Nevertheless, the inverse part of the calculations, in which the best solution is determined (minimisation), still remains a problem to be discussed -- various authors use various methods, for example, damped differential corrections \citep{pribulla2000}, the LSM \citep{panchatsaram1981}, the simplex (Levenberg-Marquart) method \citep{wade1999,lee2010}, or a combination of LSM and simplex method \citep{zasche2008}. For example, the simplex method has several versions that differ in the setting of the initial parameters, sorting conditions, or in the size of corrections. Thus, obtaining the same results, that is, repeat
the process, is very difficult, even impossible. To avoid this ambiguity, the inverse part of our code was constructed on the basis of the non-linear LSM described in \citep[e.g.,][]{mikulasek2013} applied to modelling the \lt~(for details see Sect.~\ref{modelfittingsubsec} and Appendix~\ref{liteappendix1}). This way has previously been applied to analyse the \lt~in the \object{AR~Aur} system \citep{mikulasek2011b,chrastina2013} and is similar to the one used by \citet{vanhamme2007} and \citet{wilson2014}.

\subsection{Fitting procedure}\label{modelfittingsubsec}

The code that we used is written in Matlab. It consists of several modules: loading measurements and setting initial input parameters, direct solution of the \lt~including an optional parabolic trend, inverse minimisation method, and, finally, selecting the best solution and calculating uncertainties of individual parameters through bootstrap resampling (Sect.~\ref{modelbootstrapsubsec}).  

A prediction of maxima timings $T_{\rm cal}$ calculated according to the relation
\begin{equation} 
T_{\rm cal} = M_{0} + P_{\rm puls}\times N + \mathnormal{\Delta}
\label{eq:liteEphLin}
\end{equation}
contains linear ephemeris, as well as correction $\mathnormal{\Delta}$ for the \lt. The parameter $M_{0}$ is the zero epoch of pulsations in HJD, $P_{\rm puls}$ is the pulsation period in days, and $N$ is the number of pulsation cycle from $M_{0}$.

The correction $\mathnormal{\Delta}$ for the \lt~includes calculating the orbit of a pulsating star around the centre of mass of a binary in relative units and can be expressed by the equation adopted from \citet{irwin1952a}
\begin{equation} 
\mathnormal{\Delta} = A \,\left[ (1-e^{2})\,\frac{\sin(\nu+\omega)}{1+e\,\cos\nu} + e\,\sin\omega\right],
\end{equation}\label{eq:liteDelta}
where $e$ is the numerical eccentricity, $\omega$ is the argument of periastron (usually in degrees), $A \doteq a_{1}\,\sin i/173.145$ is the projection of the semi-major axis of primary (pulsating) component $a_{1}$ in light days\footnote{the semi-amplitude of the
\lt~changes in \oc~diagram is then $A_{\rm LiTE} = A\,\sqrt{1-e^{2}\,\cos^{2}\omega}$.} according to the inclination of the orbit $i$, and $\nu$ is the true anomaly. The eccentric anomaly $E$, which is necessary to determine the true anomaly $\nu$, is solved with Kepler's equation by iteratively using Newton's method with a given precision higher than $1\times 10^{-9}\,$arcsec. Kepler's equation requires a mean anomaly $M$, which is determined from the orbital period $P_{\rm orbit}$  in days and the time of periastron passage $T_{0}$ in HJD.

Another more complex model that includes a parabolic trend in \oc~\citep[e.g.][]{zhu2012}, uses the modified Eq.~(\ref{eq:liteEphLin}) in the form of
\begin{equation} 
T_{\rm cal} = M_{0} + P_{\rm puls}\times N+ \frac{1}{2}\,P_{\rm puls}\,\dot{P}_{\rm puls}\times N^2 +\mathnormal{\Delta},
\label{eq:liteEphPar}
\end{equation} 
where $P_{\rm puls}$ is the instantaneous pulsation period at the moment $M_{0}$ and the parameter $\dot{P}_{\rm puls} = {\rm d}P_{\rm puls}/{\rm d}t$ is the rate of changes in the pulsation period in [d\,d$^{-1}$]. For easy comparison with other RR Lyrae stars, we used prescriptions from \citet{leborgne2007}. Their parameter $a_{3}=1/2\,P_{\rm puls}\,\dot{P}_{\rm puls}$ is the rate of period changes per one cycle in [d\,cycle$^{-1}$], and the rate of period changes $\beta$ in [ms\,d$^{-1}$] is $\beta = 6.31152\times10^{10}\,a_{3}/P_{\rm puls}$ or $\beta = 0.07305\times10^{10}\,a_{3}/P_{\rm puls}$ in [d\,Myr$^{-1}$] \footnote{\citet{leborgne2007} probably used a year with 366\,d, therefore their constant $0.0732\times10^{10}$ is slightly different.}. The instantaneous pulsation period at arbitrary epoch $N$ is $P_{\rm puls}(N) = P_{\rm puls}(M_{0}) + \dot{P}_{\rm puls}\times N$.

The first step of the non-linear LSM is linearising the non-linear model function (Eqs.~\ref{eq:liteEphLin}~or~\ref{eq:liteEphPar}) by Taylor decomposition of the first order \citep[see][]{mikulasek2006,mikulasek2011a}
\begin{equation} 
T_{\rm cal} \cong T_{\rm cal}(T,\textbf{b}_{0}) + \sum_{j=1}^{g} \Delta b_{j}\frac{\partial T_{\rm cal}(T,\textbf{b})}{\partial b_{j}},
\label{eq:liteEph}
\end{equation}
where $b_{j}$ are individual free parameters in vector $\textbf{b}$. Vector $\textbf{b}_{0}$ contains initial estimates of parameters, $\Delta b_{j}$ are their corrections, $g$ is the number of free parameters (the length of matrix $\textbf{b}$). 

After linearisation, the problem can be solved in the same way as in the linear LSM, but with several necessary iterations to obtain a precise solution. Our code generates the
initial parameters quasi-randomly many times from large intervals with limits that are defined by user. The derivatives are solved analytically. For more details see Appendix~\ref{liteappendix1}.

The parameter $\chi^{2}(\textbf{b}_{k})$ or its normalised value $\chi_{\rm R}^{2}(\textbf{b}_{k}) = \chi^{2}(\textbf{b}_{k})/(n-g)$, where $n$ is a number of measurements, was used as an indicator of the quality of the $k$-fit.

Since many maxima timings from the GEOS database are given without errors or are often questionable, we used an alternative approach to
determine the weights. The dataset was divided into several groups according to the type of observations (photographic, photoelectric, CCD, DSLR), and weights were assigned to each of the groups with respect to the dispersion of points around the model. The weights were improved iteratively. Groups with fewer than five points (DSLR) were merged with another group with similar data quality to avoid unrealistic weight assignment (CCD+DSLR). During the fitting process, outliers differing by more than $5\,\sigma$ from the model were rejected. All steps of the analysis were supervised visually.

The \lt~fitting process does not allow estimating the masses of the two stars, but only a mass function $f(\mathfrak{M})$
\begin{equation} 
f(\mathfrak{M}) = \frac{( \mathfrak{M}_{2}\,\sin i )^{3}}{(\mathfrak{M}_{1} + \mathfrak{M}_{2} )^2} = \frac{4\,\pi^{2}}{G}\,\frac{(a_{1}\,\sin i)^{3}}{P_{\rm orbit}^2},
\label{eq:massFunction}
\end{equation}
where $\mathfrak{M}_{1}$, $\mathfrak{M}_{2}$ are masses of the components, $i$ is the inclination angle of the orbit, $P_{\rm orbit}$ is the orbital period, and $a_{1}$ is the semi-major axis of the primary component. Based on the studies of \citet{fernley1993} and \citet{skarka2014}, we adopted as the value for the mass of the RR Lyrae component $\mathfrak{M}_{1} = 0.55$\,$\mathfrak{M}_{\odot}$ and set the inclination angle to $i = 90\,^{\circ}$ ($\sin i = 1$). This allows computing the lowest mass of the second component by solving the cubic equation
\begin{equation} 
\mathfrak{M}_{2}^3 - f(\mathfrak{M})\,\mathfrak{M}_{2}^2 - 2\,f(\mathfrak{M})\,\mathfrak{M}_{1}\,\mathfrak{M}_{2} - f(\mathfrak{M})\,\mathfrak{M}_{1}^2 = 0.
\label{eq:cubicEquation}
\end{equation}
We expect that the variation in the \oc~diagram of \object{TU~UMa} can be well described using Eq.~\ref{eq:liteEphPar} and that other possible secular variation of the pulsation period can be neglected (it has a low amplitude or appears to be on a longer timescale).

\subsection{Bootstrap-resampling}\label{modelbootstrapsubsec}

The (non-linear) LSM itself gives an estimate of the uncertainty of all fitted parameters, but the method is extremely sensitive to data characteristics. A slight change of the dataset by adding one single measurement can cause a significant difference in the new parameters from the previous solution. We therefore decided to use a statistical approach represented by bootstrap-resampling to estimate the errors.

The parameters from the best solution were used as initial values to fit a new dataset, whose points were randomly selected from the original dataset. This procedure was repeated 5000 times. From the scattering of individual parameters of these five thousand solutions, we estimated their uncertainties. The errors in Tables~\ref{CLAurtable}
and~\ref{TuUMatable} correspond to 1\,$\sigma$.  

\subsection{Test object CL Aurigae: an eclipsing binary with probable \lt~and mass transfer}\label{CLAursubsec}

The code was, among others, tested on the well-known detached binary system \object{CL Aur}. This eclipsing binary was chosen because it shows
the \lt~and a secular period change. In addition, \object{CL Aur} was  studied in a similar way three times during the past 15 years \citep{wolf1999,wolf2007,lee2010}.

\begin{figure}[t]
\centering
\includegraphics[width=1.00\hsize]{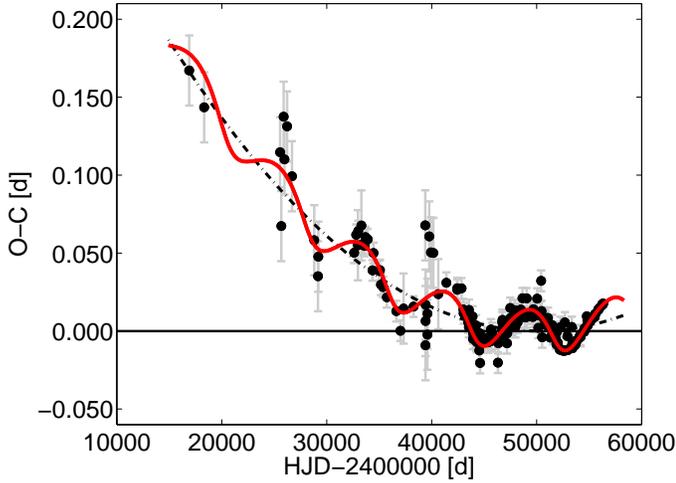}
\caption{\oc~diagram of the testing eclipsing binary CL Aurigae (double star without an RR Lyrae component) with the \lt~and parabolic trend (black circles). The model of changes (red line) is based on our parameters from Table~\ref{CLAurtable}.}
\label{Fig:CLAur}
\end{figure}

The times of minima of \object{CL Aur} taken from the \oc~gateway database\footnote{http://var.astro.cz/ocgate/, all used minima timings are available at the CDS portal.} \citep{paschke2006} were used to construct \oc~diagram and to determine parameters through the methods described above\footnote{The model was calculated for the half-value of the period. Subsequently, results were corrected for this effect.}. Our best model (Fig.~\ref{Fig:CLAur}) describes the \oc~variations very well in the most recent part (precise CCD observations). The old part of the \oc~diagram with visual and photographic measurements is highly scattered, but these measurements were also taken into account during the fitting process by assigning them a lower weight (model weights for different observation methods were found in the ratio pg:vis:ccd 1:11.5:403)\footnote{\citet{wolf2007} visually distinguished the data quality by weights in each category 0, 1, 2 for pg, 0, 1, 2 for visual, and 5, 10, 20 for CCD observations, respectively.}. We conclude that our results are comparable with previous results (Table~\ref{CLAurtable}). This example, as well as additional testing, showed that our code works well and is suitable for analysing RR Lyraes with suspected \lt. The code was successfully used to model the \lt~in \object{V2294 Cyg} \citep{liska2014b}.

\begin{center}
\begin{table}[t]
\caption{Our determined parameters from the testing object \object{CL~Aur} (right) together with results from previous studies (left).}\label{CLAurtable}
{\tiny 
\begin{center}
\def\arraystretch{1.5}
\tabcolsep=1.1pt
\begin{tabular}{lcc|c}
\hline\hline
Study                                                   & \citet{wolf2007}                         & \citet{lee2010}                               & Our model\\
\hline
$P(M_{0})$ [d]                          & 1.24437505(18)                                & 1.24437498(17)                          & 1.24437488$^{+16}_{-12}$\\
$M_{0}$ [HJD]                                   & 2450097.2712(5)                               & 2450097.27082(46)                               & 2450097.2716$^{+6}_{-7}$\\
$10^{-10}$$\dot{P}$                             & \multirow{2}*{$4.05(6)^{*}$}  & \multirow{2}*{$3.92(55)^{*}$}   & \multirow{2}*{3.76$^{+30}_{-25}$}\\                   
$$[d\,d$^{-1}$]                                 &                                                               &                                                                 &\\
$10^{-10}\,$$a_{3}$                             & \multirow{2}*{$2.52(4)$}              & \multirow{2}*{$2.44(34)$}               & \multirow{2}*{2.34$^{+17}_{-14}$}\\   
$$[d\,cycle$^{-1}$]                     &                                                               &                                                                 &\\
$\beta$ [ms\,yr$^{-1}$]                 & $12.8(2)^{*}$                                 & $12.4(1.7)^{*}$                         & 11.9$^{+9}_{-7}$\\                    
$\beta$ [d\,Myr$^{-1}$]                 & $0.148(2)$                                    & $0.143(20)$                                     & 0.137$^{+10}_{-8}$\\  
$P_{3}$ [yr]                                    & 21.7(2)                                               & 21.63(14)                                               & 21.61$^{+19}_{-18}$\\
$T_{0}$  [HJD]                                  & 2443880(80)                                   & 2444072(56)                                     & 2444020$^{+140}_{-190}$\\
$e$                                                             & 0.32(2)                                         & 0.337(53)                                             & 0.27$^{+5}_{-3}$\\
$\omega$ $[^{\circ}]$                   & 209.2(1.2)                                    & 218.9(2.7)                                      & 218$^{+6}_{-9}$\\
$A$     [light day]                                     & 0.0144(12)$^{*}$                              & 0.01378(72)$^{*}$                               & 0.01388$^{+35}_{-25}$\\
$a_{12}\sin i$ [au]                             & 2.49(22)$^{*}$                                & 2.38(12)                                                & 2.40$^{+6}_{-4}$\\
$f(\mathfrak{M}_{3})$ [$\mathfrak{M}_{\odot}$]          & 0.034                                                 & 0.0290(15)                                      & 0.0297$^{+19}_{-14}$\\
$K_{12}$ [km\,s$^{-1}$]                 & $-$                                                   & $-$                                                     & 3.44$^{+10}_{-6}$\\   
$\chi_{\rm R}^2$                                & $-$                                                   & $-$                                                     & 1.04(10)\\
$N_{\rm min}$                                   & 144                                                   & 198                                                     & 203\\
\hline
\end{tabular}

\end{center}
{\bf Notes.} $^{(*)}$ Parameter calculated using values from the original study.}
\end{table}
\end{center}

\section{LiTE in \object{TU UMa} and analysis of the $\textrm{\textit{O-C}}$~diagram}\label{resultstuumasubsec}
Cyclic changes in \oc~diagram of \object{TU UMa} have been well known for a long time and were analysed in detail by \citet{saha1990a}\footnote{Several parameters from \citet{saha1990a} e.g. $A$, $a\,\sin i$ were corrected in their erratum \citep{saha1990b}.}, \citet{kiss1995}, and \citet{wade1999}. The parameters determined by these authors are given in Table~\ref{TuUMatable}.

\begin{figure}[t]
\centering
\includegraphics[width=1.00\hsize]{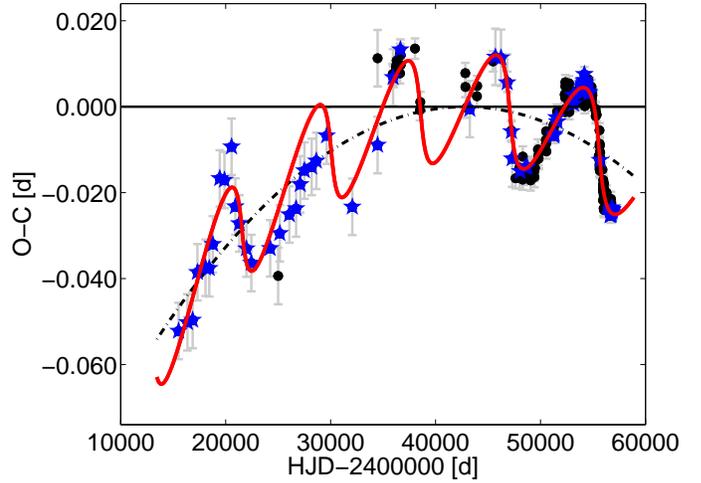}
\includegraphics[width=1.00\hsize]{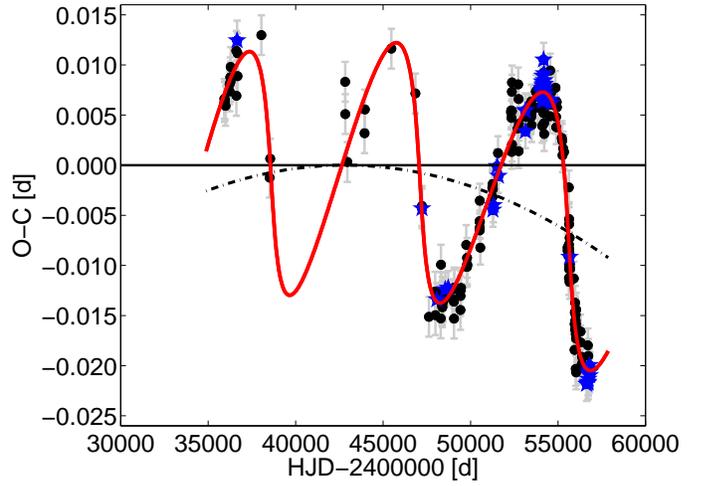}
\caption{\oc~diagram of \object{TU UMa}. Black circles and blue stars display the maxima adopted from the GEOS database
and new maxima determined in this work. The period decrease manifested by the parabolic trend (dot--dashed line) is obvious. Cyclic changes due to an orbital motion are also clearly visible. Our model of LiTE is represented by the solid red line. The top panel shows model 1 with all available data, while the plot in the bottom panel shows the situation with only photoelectric, CCD, and DSLR observations (model 2).}
\label{Fig:TUUMa}
\end{figure}

Our dataset described in Sect.~\ref{datasec} is much denser and more extended than in previous studies (in contrast to \citet{wade1999}, who used only 83, we used 253
maxima timings). Our \oc~values span 113 years, which is mainly due to measurements recorded on the Harvard plates (provided by the project DASCH) in the first half of the twentieth century. Because the complete dataset is not homogeneous (it has incomparably better coverage over the last two decades with CCD measurements), we analysed \lt~in \object{TU UMa} in two ways. Model 1 is based on the whole dataset\footnote{Weights for model 1 were found to be of the ratio pg\,:\,pe\,:\,CCD+DSLR 1.0\,:\,7.6\,:\,14.8, uncertainties are 0.0066\,d, 0.0024\,d, and 0.0017\,d.}, while model 2 describes only photoelectric, CCD, and DSLR observations\footnote{Weights for model 2 were of the ratio pe\,:\,CCD+DSLR 1.00\,:\,1.38, uncertainties are 0.0020\,d and 0.0017\,d.}. Since \object{TU UMa} experiences secular period decrease (in Fig.~\ref{Fig:TUUMa} represented by the parabolic dot-dashed curve) with the rate $1/2\,P_{\rm puls}\,\dot{P}_{\rm puls}$ of about $-2.9\times10^{-11}$ days per cycle \citep{wade1999}, we used the complex form of the model (Eq. \ref{eq:liteEphPar}).

\begin{figure}[t]
\centering
\includegraphics[width=1.00\hsize]{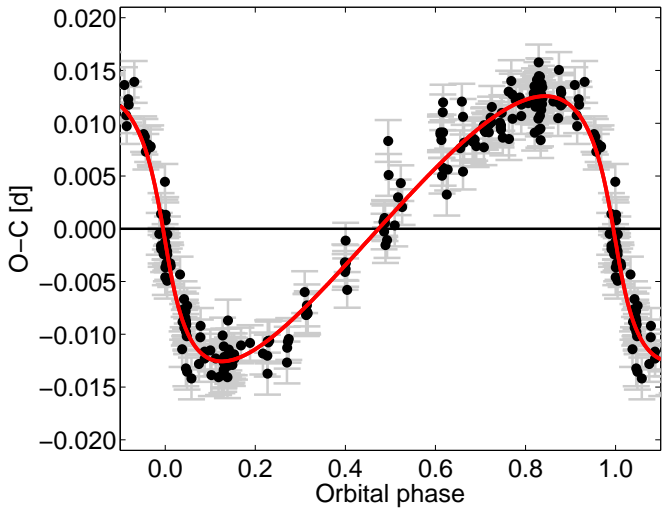}
\caption{\oc~diagram of \object{TU UMa} constructed only from photoelectric, CCD, and DSLR measurements after subtracting the parabolic trend and phased with the orbital period based on model~2.}
\label{Fig:TUUMaPhase}
\end{figure}

Logically, model 1, which is based on the whole data set (covering five orbital periods) gives more precise, but slightly different results than model 2, which spans only two of the 23-year orbital cycles. Nevertheless, high-accurate photoelectric and CCD measurements cover the whole orbital cycle very well (see Fig.~\ref{Fig:TUUMaPhase}). Because of the shorter time base, our second model has a lower value of the secular period evolution, for example. This happens at the expense of an increasing eccentricity (0.66 and 0.69 for models 1 and 2, respectively) and change in the argument of periastron (181 and 184$^{\circ}$).

\begin{center}
\begin{table*}[t]
\caption{Parameters determined previously (left) and our results (right) for the system \object{TU UMa}. The mass limit of the second body was estimated from the mass function $f(\mathfrak{M})$, the orbit
inclination $(i=90^{\circ}),$ and the mass of the RR Lyrae star $\mathfrak{M}_{1} = 0.55$\,$\mathfrak{M}_{\odot}$ adopted from \citet{fernley1993} and \citet{skarka2014}. Our parameters (right part of the table) were calculated from all maxima timings for model~1 and only from photoelectric, CCD, and DSLR measurements for model~2. }\label{TuUMatable}
{\tiny 
\begin{center}
\def\arraystretch{1.3}
\begin{tabular}{lccc|cc}
\hline\hline
Study                                           & \citet{saha1990a}             & \citet{kiss1995}                & \citet{wade1999}$^{C}$        & model 1                                         &  model 2\\
\hline
$P_{\rm puls}(M_{0})$ [d]                       & 0.5576581097                 & 0.5576581097                  & 0.55765817(29)$^{D}$           & 0.557657598$^{+19}_{-20}$             & 0.557657477$^{+20}_{-25}$\\
$M_{0}$ [HJD]                           & 2425760.4364          & 2425760.4364                    & 2425760.464(5)$^{D}$          & 2442831.4869$^{+4}_{-5}$                & 2442831.4864$^{+4}_{-4}$\\
$\dot{P}_{\rm puls}$            & \multirow{2}*{$\times$}       & \multirow{2}*{$-31.48^{*}$}     &  \multirow{2}*{$-10.4(7)^{*}$}                &  \multirow{2}*{$-6.99^{+30}_{-25}$}&  \multirow{2}*{$-4.55\,^{+50}_{-40}$}\\                    
$10^{-11}$[d\,d$^{-1}$] &       &  & &\\
$a_{3}\!=\!1/2\,P_{\rm puls}\dot{P}_{\rm puls}$         & \multirow{2}*{$\times$}       & \multirow{2}*{$-8.78^{*}$}      &  \multirow{2}*{$-2.9(2)$}             &  \multirow{2}*{$-1.95^{+8}_{-7}$}&  \multirow{2}*{$-1.27\,^{+13}_{-10}$}\\              
$10^{-11}\,$[d\,cycle$^{-1}$] & &  & &\\
$\beta\!=\!\dot{P}_{\rm puls}$ [ms\,yr$^{-1}$]          & $\times$      & $-9.934^{*}$                            & $-3.3(2)^{*}$                                 & $-2.21^{+9}_{-7}$                       & $-1.44\,^{+15}_{-11}$\\                       
$\beta\!=\!\dot{P}_{\rm puls}$ [d\,Myr$^{-1}$]          & $\times$      & $-0.11498^{*}$                  & $-0.038(3)^{*}$                               & $-0.0255^{+10}_{-8}$                    & $-0.0166\,^{+17}_{-13}$\\                     
$P_{\rm orbit}$ [yr]            & 20.19$^{*}$                   & 24.1(3)$^{*}$                   & 23.27(24)$^{*}$                       & 23.30$^{+6}_{-6}$                       & 23.27$^{+6}_{-8}$\\
$T_{0}$  [HJD]                          & 2425000$^{*}$                 & 2447200(50)                     & 2421585(207)                          & 2447092$^{+40}_{-40}$                   & 2447124$^{+45}_{-40}$\\
$e$                                                     & 0.970                                 & 0.90(5)                         & 0.74(10)                                      & 0.663$^{+30}_{-35}$                     & 0.686$^{+25}_{-25}$\\
$\omega$ $[^{\circ}]$           & 196.1$^{*}$                   & 178(3)                          & 183(5)$^{*}$                          & 181.3$^{+2.5}_{-2.0}$                   & 184.1$^{+2.0}_{-2.0}$\\
$A$     [light day]                             & 0.056                                 & 0.023(5)$^{*}$          & 0.0203(35)$^{*}$                      & 0.0168$^{+5}_{-6}$                      & 0.0172$^{+6}_{-5}$\\
$a_{1}\sin i$ [au]                      & 10                    & 4.0(7)$^{*}$                    & 3.52(61)                                      & 2.91$^{+9}_{-10}$                       & 2.99$^{+11}_{-9}$\\
$f(\mathfrak{M})$ [$\mathfrak{M}_{\odot}$]              & $-$                                   & 0.11(1)                         & 0.080                                         & 0.046$^{+5}_{-4}$                       & 0.049$^{+6}_{-4}$\\
$\mathfrak{M}_{\rm 2, min}$\,$^{*}$ [$\mathfrak{M}_{\odot}$]    & $-$                     & 0.17                                  & $-$                                             & 0.327$^{+14}_{-14}$                   & 0.339$^{+16}_{-14}$\\
$K_{1}$ [km\,s$^{-1}$]          & 60.7$^{*}$                                    & 11.4(5)                         & 6.6                                           & 4.97$^{+35}_{-35}$                      & 5.25$^{+40}_{-30}$\\  
$\chi_{\rm R}^2$                        & $-$                                   & $-$                                     & $-$                                           & 1.05(9)                                         & 1.03(10)\\
$N_{\rm max}$                           & $\sim$\,43                    & $\sim$\,42                      & 67                                            & 253                                                     & 220\\[1mm]
\hline
\end{tabular}
\end{center}
{\bf Notes.} $^{(*)}$ Parameter calculated using values from the original study, $^{(C)}$ their approach C was selected, $^{(D)}$ pulsation elements are known only from their approach D.}
\end{table*}
\end{center}

In comparison with orbital elements from previous studies given in Table~\ref{TuUMatable}, our results have a higher confidence level that is due to the larger and better dataset. Our values differ mainly in eccentricity and distance between components and in the mass function. All these values were found to be significantly lower than those from \citet{saha1990a}, \citet{kiss1995}, or \citet{wade1999}. Values from \citet{saha1990a} differ more because
they neglected the period decrease.

The eccentricity is not as extreme as proposed by \citet{saha1990a}. Its value is comparable with other systems from \citet{hajdu2015,li2014}.

 Based on our results, it seems that \object{TU UMa} is very likely a member of a well-detached system with a dwarf component with a minimal mass of only 0.33\,$\mathfrak{M}_{\odot}$. Since no signs of the companion are observed in the light of \object{TU UMa}, it is probably a late-type main-sequence dwarf star. However, we cannot exclude the possibility that it is a white dwarf or a neutron star, since we do not know the inclination.

Except for the \lt, which is the most remarkable feature of the \oc~diagram, changes represented by the parabolic trend are also
apparent. The progression of the dependence suggests secular shortening of the pulsation period of \object{TU UMa}, which is almost certainly an evolutionary effect because the mass transfer, which is responsible for period changes in close binaries, can be excluded because of the very wide orbit of \object{TU UMa}. In addition, value $\beta = \dot{P}_{\rm puls} \!\sim\! -2.2$\,ms\,yr$^{-1}\!=\!-0.026$\,d\,Myr$^{-1}$ can correspond to a blueward evolution of the RR Lyrae component, but \citet{leborgne2007}, for instance, reported a higher median value $\beta = -0.20$\,d\,Myr$^{-1}$ for their sample of 21 stars with a significant period decrease.

\begin{figure}[t]
\centering
\includegraphics[width=1.00\hsize]{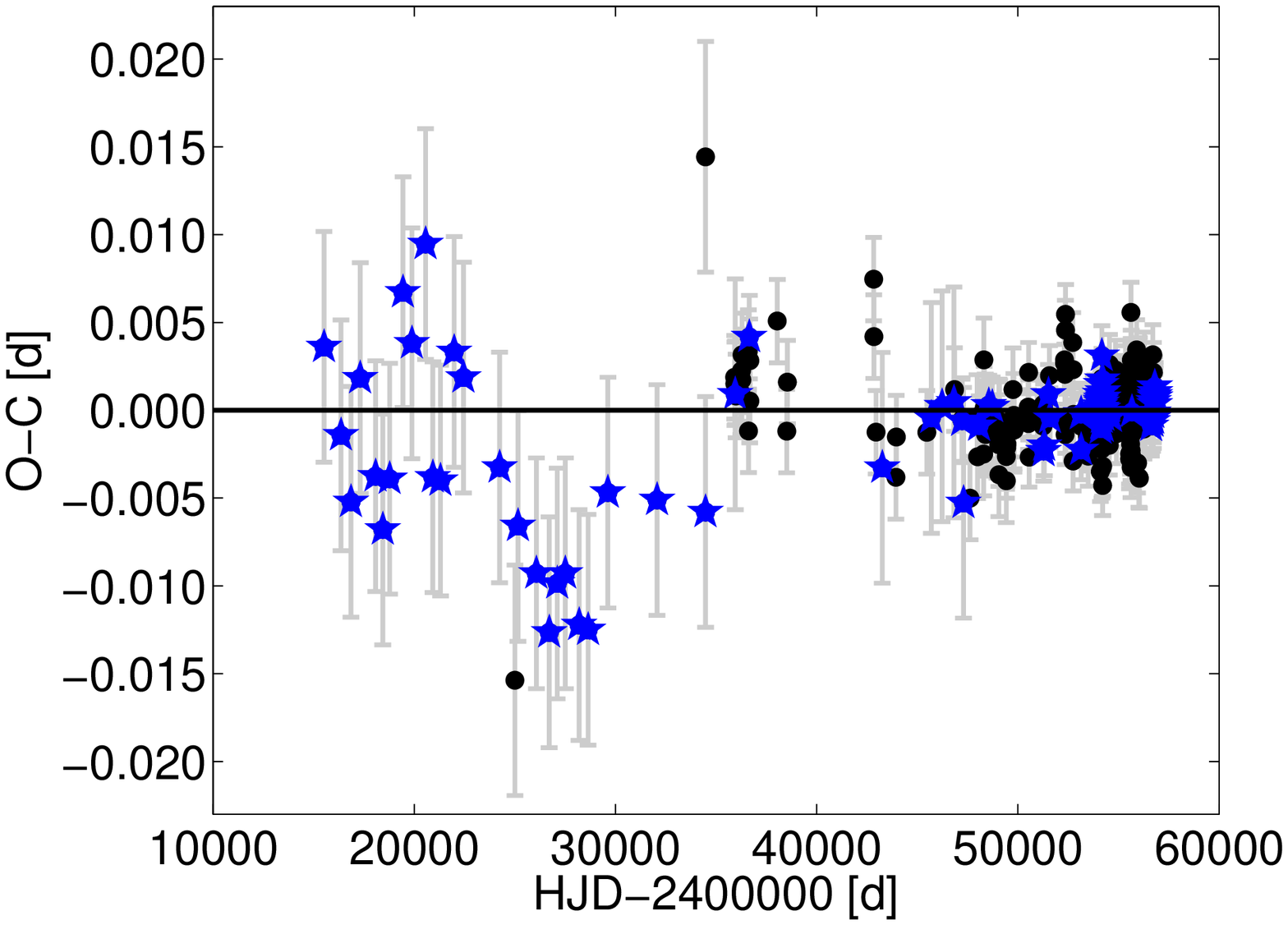}
\includegraphics[width=1.00\hsize]{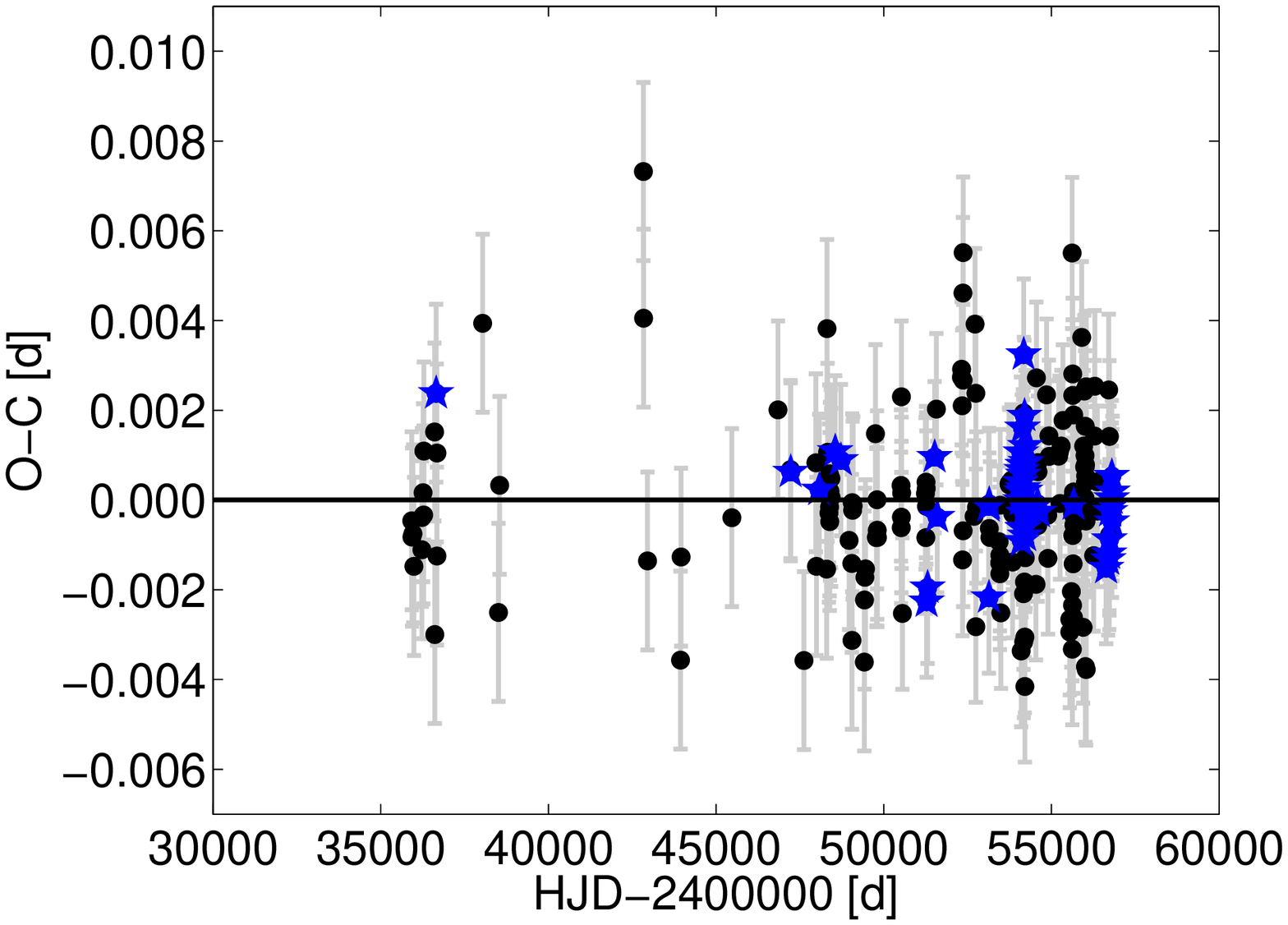}
\caption{Residual \oc~diagram of \object{TU UMa} after subtracting the first LiTE model~(top panel) and second model -- only photoelectric, CCD, and DSLR observations (bottom). Black circles and blue stars display maxima adopted from the GEOS database
and new maxima determined in this work. The jump in the general trend of \oc~in the range from JD 2425000 to 2429000 (top panel) might be an indication of a more complex period change.}
\label{Fig:ocResidual}
\end{figure}

After subtracting our model 1 from the whole dataset~(Fig.~\ref{Fig:ocResidual}), several photographic measurements (in the range of between JD 2425000 and 2429000) are deviate more than other values, with a systematic shift of about 15 minutes (0.01\,d). This might indicate that \object{TU UMa} undergoes more complex period changes than the \lt~and parabolic trend alone. Some possible explanations such as cubic trend or an additional \lt~could describe this variation. However, even though the influence of longer exposure of photographic measurements was taken into account, some other instrumental artefact might play a role.

\section{Other proofs for binarity}\label{proofsec}

Since the \lt~is only an indirect manifestation of binarity, it is necessary to prove it in a different way. The analysis of the mean RVs can be considered as the most valuable test. In addition to this method, other possible approaches with which
the binarity of \object{TU UMa} can be confirmed are discussed in the next sections.

\subsection{Radial velocities}\label{rvsec} 

The known orbital parameters from the analysis of the \lt~allow us to predict, but also reconstruct, the RV curve from the past. The binarity can then be proved by comparing the model for the orbital RV curve and the spectroscopically determined centre-of-mass RV. This analysis was first performed for \object{TU UMa} by \citet{saha1990a} and a few years later by \citet{wade1999}. They noted systematic shifts in the RV determined at different times. Their predictions correlate fairly well with measurements.

We scanned the literature for RV measurements and found nine sources (Table~\ref{RVPublicationtable}). Unfortunately, the last available dataset with RVs was published in 1997. \citet{saha1990a} used only three RV sources, \citet{wade1999} did not give their values. In addition, both authors ignored the RV measurements from \citet{preston1964}. Other authors have reported slightly different values determined from the same dataset (Table~\ref{Tab:RVvalues}, Col. 2) without listing the mean time of observation. Therefore we decided to re-analyse all available RV measurements.

\begin{figure}[t]
\centering
\includegraphics[width=1.00\hsize]{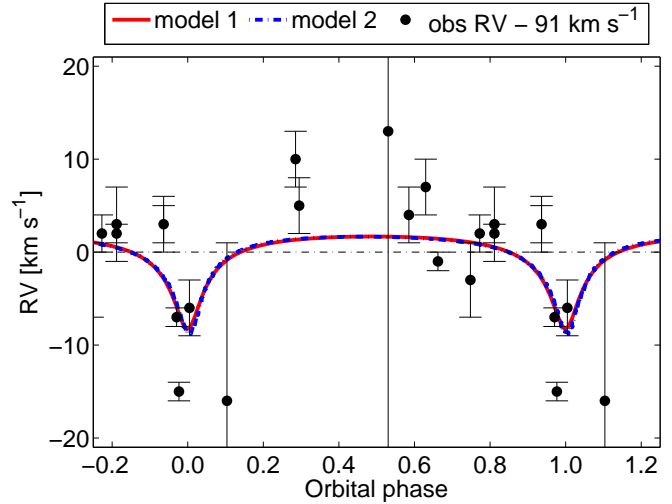}
\caption{Models of variations in RV caused by orbit of pulsating component around mass-centre of the binary system (red and blue lines) and center-of-mass velocities determined for each dataset of RV measurements using template fitting or adopted from literature. The visually estimated correction $-91$\,km\,s$^{-1}$ for systematic velocity mass-centre of the system from Sun ($\gamma$-velocity) was applied.}
\label{Fig:TUUMaRVmodel}
\end{figure}

Determining the centre of mass of the RV for a binary with a pulsating star is more complicated than for a binary with non-variable stars (pulsations are often the dominant source of RV changes). Other inconveniences are connected with an RV based on different types of spectral lines (e.g. Balmer or metallic lines). They are formed at different depths, and therefore RV curves from different lines have different shapes, amplitudes, and zero points \citep[shown e.g. by][]{sanford1949,oke1962}. Since available RV measurements are based on various lines, it was necessary to unify them. This was done using the highly accurate normalised template curves from \citet{sesar2012}. Firstly we modelled these template curves with an $n$-order harmonic polynomial. The observed RV curve for the particular spectral line was then compared with the polynomial image of the template, and the amplitude and the central value of RV curve was simultaneously determined by the LSM for all datasets. Before this step, measurements were time-corrected for binary orbit and period shortening (based on model 1), and several of the datasets were divided into smaller groups to obtain a time resolution of about one year (see Table \ref{Tab:RVvalues} with the determined mean RVs for given epochs corresponding to the mean value of observation time). We did not find original RV measurements for two studies \citep{fernley1997,solano1997} and therefore adopted their mean RV values and estimated the
mean time of observation from the information provided in their papers. Finally, the mean centre of mass of the RV values were compared with the RV model resulting from our \lt~analysis (Fig.~\ref{Fig:TUUMaRVmodel}). The points roughly follow the model RV curve. 

\begin{center}
\begin{table}[t]
\centering
\caption{Sources of RV measurements for \object{TU UMa}, $S$ is a source number, $N_{\rm RV}$ is a number or RV measurements.}\label{RVPublicationtable}
{\tiny
\def\arraystretch{1.5}
\tabcolsep=1.3pt
\begin{tabular}{llcc}
\hline\hline
$S$                     & Publication                                                   & $N_{\rm RV}$    & Lines\\
\hline
{\bf 1}         & \citet{abt1970}                                               & 1                               & Unknown\\
{\bf 2}         & \citet{barnes1988}                                    & 74                      & Metallic\\
{\bf 3}         & \citet{fernley1997}                                   & 3?                      & OI triplet, (H$_{\alpha}$)\\
{\bf 4}         & \citet{layden1993,layden1994}                 & 5                               & Hydrogen, CaII K\\
{\bf 5}         & \citet{liu1989}                                               & 60                      & Metallic\\
{\bf 6a}        & \multirow{2}*{\citet{preston1961}}    & 4                             & Metallic\\
{\bf 6b}        &                                                                               & 21                      & H$_{\gamma}$, H$_{\delta}$, H$_{8-11}$, Ca II K\\
{\bf 7a}        & \multirow{2}*{\citet{preston1964}}    & 12                    & Metallic\\
{\bf 7b}        &                                                                               & 8+7                     & Hydrogen\\
{\bf 8}         & \citet{saha1990a}                                             & 32                      & Metallic\\
{\bf 9}         & \citet{solano1997}                                    & 3?                      & Metallic, (H$_{\gamma}$)\\
\hline
\end{tabular}}
\end{table}
\end{center}

\begin{center}
\begin{table}[t]
\begin{center}
\caption{Determined values of centre-of-mass velocities for \object{TU UMa} based on different RV measurements and templates from \citet{sesar2012}. The mean values published in different publications are present for comparison, $S$ is the source number of the original RV measurements from Table~\ref{RVPublicationtable}.}\label{Tab:RVvalues}
{\tiny
\def\arraystretch{1.5}
\tabcolsep=1.3pt
\begin{tabular}{lc|ccc}
\hline\hline
$S$                     & RV$_{\rm pub}$        & $T_{\rm mid}$ & RV$_{\rm our}$  & errRV$_{\rm our}$     \\      
                        & [km\,s$^{-1}$]        & [HJD]                 & [km\,s$^{-1}$]  & [km\,s$^{-1}$]        \\      
\hline
{\bf 1}         & 104(35)$^{La}$        & 2426076               & 104$^{La}$              & 35$^{La}$ \\                                  
\hline
{\bf 2}         & 90(2)$^{F}$, 90(2)$^{So}$                                     & 2443563 & 95            & 3     \\                                      
                        &                                       & 2443941         & 98                            & 3     \\                                      
                        &                                       & 2444218         & 90                            & 1     \\                                      
                        &                                       & 2444948         & 88                            & 4     \\                                      
\hline
{\bf 3}         & 101(3)$^{F}$, 101(5)$^{So}$                           & 2449520 & 101$^{F}$                             & 3$^{F}$       \\                                      
\hline
{\bf 4}         & 75(17)$^{La}$, 75(17)$^{F}$                           & 2447975 & 75$^{La}$                             & 17$^{La}$     \\                                      
\hline
{\bf 5}         & 84.2$^{Li}$, 84$^{Sa}$, 84(1)$^{La}$,         & 2446843 & 84                            & 1     \\                                      
                        & 84(1)$^{F}$, 84(2)$^{So}$                                     & 2447130 & 85                            & 3     \\                                      
\hline
{\bf 6a}        & --                            & 2436979               & 93                              & 1     \\                                      
\hline
{\bf 6b}        & 92(1)$^{P}$, 87$^{H}$,                                        & 2436648 & 93                            & 2     \\                                      
                        & 93(3)$^{Sa}$, 92(1)$^{F}$                                     & 2436979 & 94                            & 4     \\                                      
\hline
{\bf 7a}        & --                            & 2438039               & 94                              & 2     \\                                      
\hline
{\bf 7b}        & --                            & 2438039               & 94                              & 3     \\                                      
\hline
{\bf 8}         & 77$^{Sa}$, 77(2)$^{F}$, 77(2)$^{So}$          & 2446894 & 76                            & 1     \\                                      
\hline
{\bf 9}         & 96(3)$^{So}$          & 2449600               & 96$^{So}$                       & 3$^{So}$      \\                                      
\hline
\end{tabular}}
\end{center}
{\bf Notes.} Values adopted from $^{(F)}$\citet{fernley1997}, $^{(H)}$\citet{hemenway1975}, $^{(La)}$\citet{layden1994}, $^{(Li)}$\citet{liu1990}, 
$^{(P)}$\citet{preston1961}, $^{(Sa)}$\citet{saha1990a}, $^{(So)}$\citet{solano1997}.
\end{table}
\end{center}

\begin{figure}[t]
\centering
\includegraphics[width=1.00\hsize]{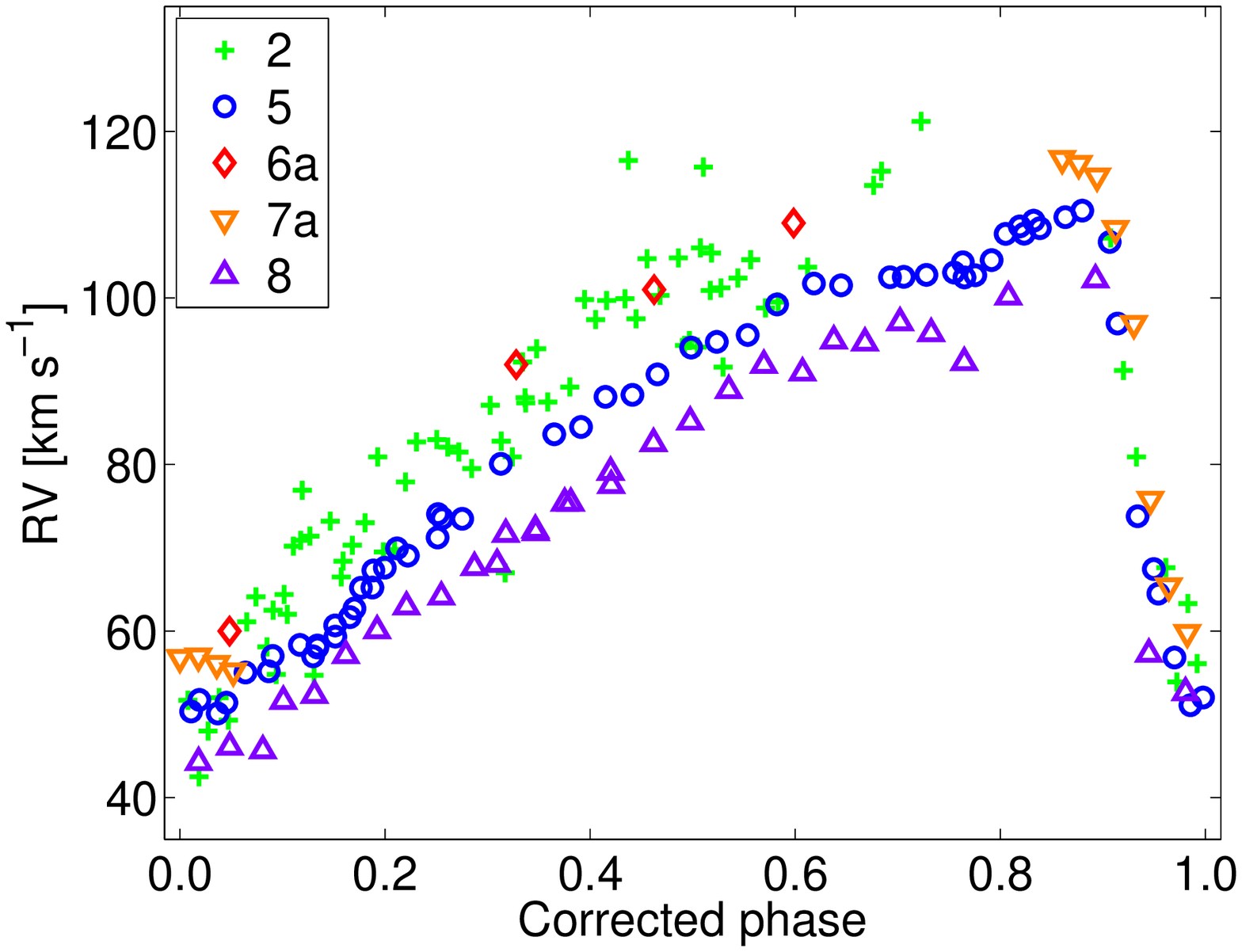}
\includegraphics[width=1.00\hsize]{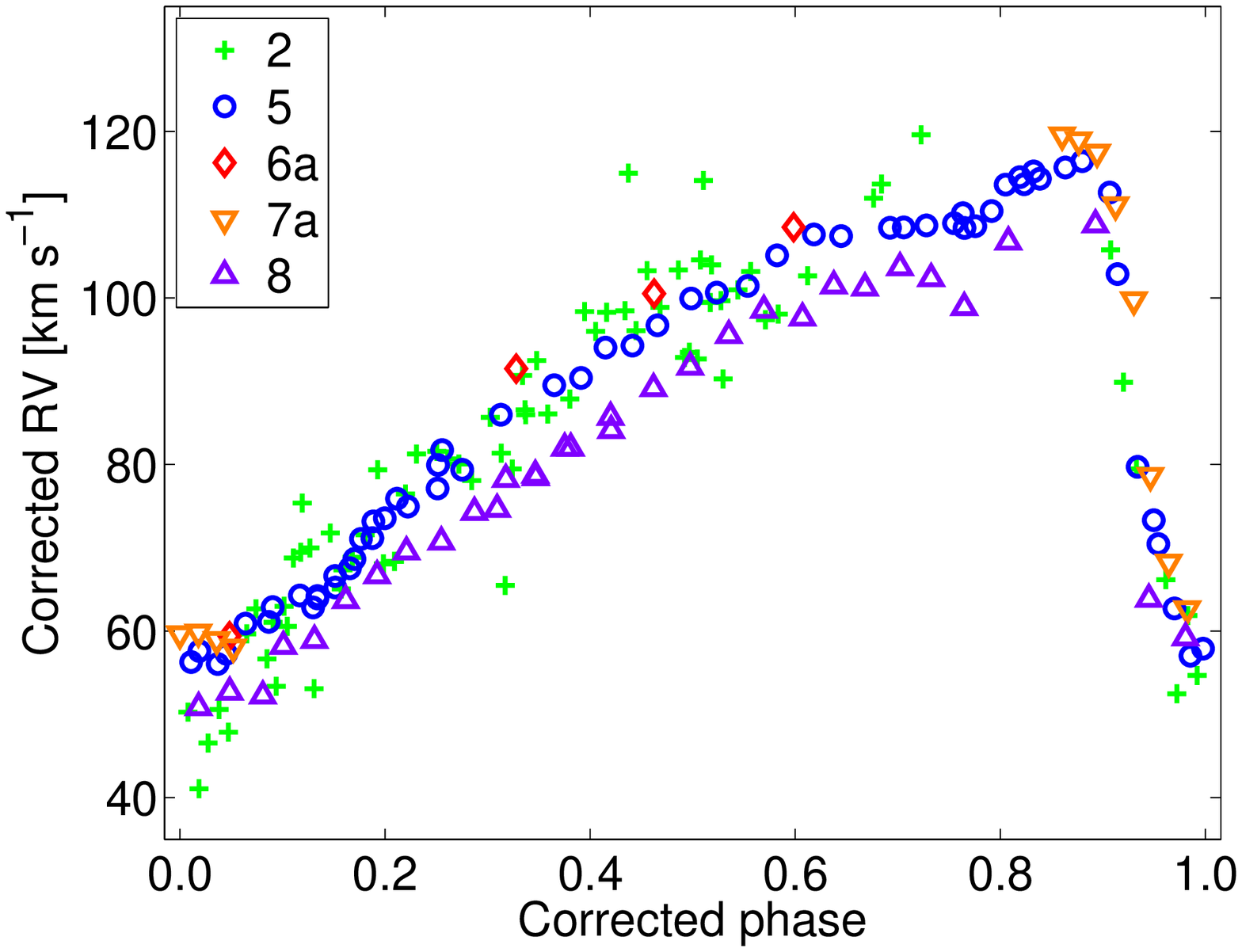}
\caption{Radial velocity curves from the metallic lines of \object{TU UMa} from different publications phased with the pulsation period corrected for the \lt~and secular period changes. Uncorrected observed RVs (top) and corrected values after subtracting the
changes in RV caused by orbital motion based on our model 1 (bottom). RV values corrected for binary orbit are evidently less scattered than uncorrected RVs.}
\label{Fig:TUUMaRVphase}
\end{figure}

An alternative test for binarity using RV curves can be performed by comparing the observed RV curves (the top panel of Fig.~\ref{Fig:TUUMaRVphase}) and those in which the orbital RV curve from the model is subtracted (the bottom panel of Fig.~\ref{Fig:TUUMaRVphase}). In~this figure RV measurements are phased according to the pulsation period\footnote{The stitching in phase was possible only by taking the \lt~and secular period change into account. Without this, the curve was scattered horizontally.}. Apparently, the phased RV curve with corrected velocities is significantly less vertically scattered than without the correction. The residual scatter in the bottom panel results from the different metallic lines that the RVs were based on.

Both tests clearly show that \object{TU UMa} is very likely bound in a binary system.

\subsection{Eclipses}\label{eclipsesec}
A detection of eclipses in the light curve (in the appropriate phase of the orbit) would be a strong proof for binarity of \object{TU UMa}. Several third components of eclipsing binary stars that
were known only from the \lt\ were confirmed by detecting additional eclipses \citep[e.g. in the {\it Kepler} project,][]{slawson2011}. The probability of catching an eclipse in \object{TU~UMa} is very low because the expected orbital period of the binary system is very long and radius and luminosity of the secondary component are probably much smaller and lower than for the pulsating star. In addition, the inclination angle of the orbit is unknown. Our two preliminary LiTE models allow us to estimate the time of a possible eclipse, where the RR Lyrae component should transit the secondary one (January -- February 2014 or May -- June 2014), but the difference between the
two predictions is too large. However, we attempted to detect the proposed eclipse.

\begin{figure}[t]
\centering
\includegraphics[width=1.00\hsize]{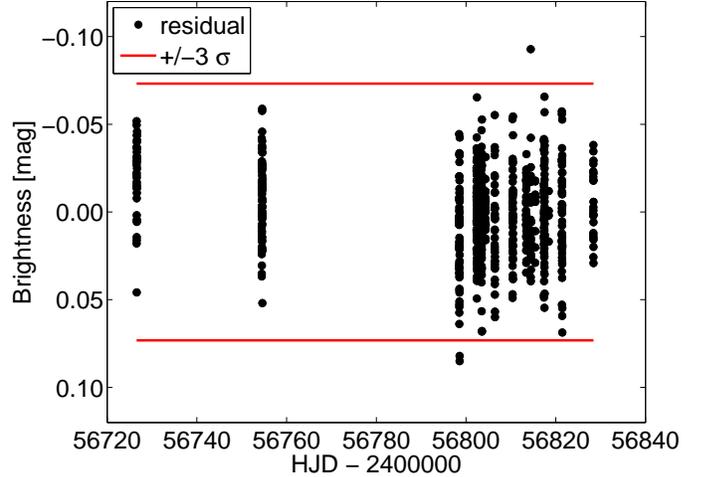}
\caption{Residuum of the light curve of \object{TU UMa} after subtracting
the harmonic polynomial model. No signs of an eclipse with an
amplitude higher than 0.07\,mag was detected in the \textit{green} band.}
\label{Fig:LightCurveResidual}
\end{figure}

Observations with the small telescope described in Sect.~\ref{observsubsec} were dedicated for this purpose. Unfortunately, our measurements were insufficient for a reliable decision about eclipses. Weather conditions, limited object visibility and other influences allowed
us to observe in only 19 nights, which is hardly sufficient considering the imprecise eclipse prediction. At least we can conclude that no sign of an eclipse with an amplitude higher than 0.07\,mag was detected in our data (see Fig.~\ref{Fig:LightCurveResidual}).

\section{Summary and conclusions}\label{summarysec}

We presented a new analysis of a probable \lt~in \object{TU~UMa}. We used published maxima timings from the GEOS database (168 values) and added maxima values from our photometric observations and from the SuperWASP and Pi of the Sky surveys. We applied the template fitting method to determine maxima from these measurements and also from sparse data from the projects Hipparcos, NSVS and DASCH. Altogether, we analysed 253 maxima timing measurements, which is about three times more than were used in the dataset in the last study of \object{TU UMa} by \citet{wade1999}. This large and well covered dataset allowed us to determine a quadratic ephemeris of the pulsations and orbital elements of the binary system with much better accuracy than in previous studies (Table~\ref{TuUMatable}). All analyses were performed with a new code written in Matlab that uses a bootstrap method to estimate the errors. We calculated two models: model~1, which describes the whole dataset (without visual maxima timings), and model~2, which describes only high-accurate photoelectric, CCD and DSLR maxima. The second model is based on data with a significantly shorter time span than for model 1.

The second model gives a lower value of the period-decrease rate ($\beta=\dot{P}_{\rm puls} \!\sim\! -1.4$\,ms\,yr$^{-1}$), which causes the eccentricity to become higher ($e \!\sim\! 0.69$) than in the first model ($\beta \!\sim\! -2.2$\,ms\,yr$^{-1}$, $e \!\sim\! 0.66$). For comparison, \citet{arellanoferro2013} give $\beta \!=\! -1.3$\,ms\,yr$^{-1}$ without fitting the \lt. Nevertheless, both our models have a lower eccentricity value, semi-major axis of pulsating component $a_{1}\,\sin i$ (2.9\,au or 3.0\,au), and semi-amplitude of RV variations of the pulsating star $K_{1}$ (5.0\,km\,s$^{-1}$ or 5.3\,km\,s$^{-1}$) than in previous works. Our values of the orbital period (identically 23.3\,yr) and argument of periastron $\omega$ (181$^{\circ}$ or 184$^{\circ}$) are comparable with values determined by previous authors. In addition, the lowest mass limit of the secondary component (0.33\,$\mathfrak{M}_{\odot}$ or 0.34\,$\mathfrak{M}_{\odot}$) was determined with an assumption for the mass of the RR Lyrae component of 0.55\,$\mathfrak{M}_{\odot}$.

The binary nature was tested in several ways. Firstly, our models of the orbit gave predictions of possible eclipses. Although the prediction was highly inaccurate and an eclipse is highly unlikely (wide orbit, unknown inclination, and other important parameters) we attempted to detect them, but were unfortunately not successful. 

Binarity is expected manifest itself in cyclic changes of the mean RV. We adopted RV measurements from nine independent sources and corrected their values according to our model by subtracting the \lt~and secular changes. When an observed RV curve was phased with the pulsation period, we obtained the typical RV curve for RR Lyrae, which was scattered. The scatter significantly decreased when our model was applied.

We also determined central RV values for each RV dataset using pulsation templates for different spectral lines from \citet{sesar2012}. We compared these values with our model of orbital RV variations based on orbital parameters known from the \lt. The correlation is evident (Fig.~\ref{Fig:TUUMaRVmodel}). 

The two successful proofs are important for confirming the binarity of \object{TU UMa}. However, only long-term spectroscopic measurements covering the whole orbital cycle can unambiguously confirm that \object{TU~UMa} is indeed a member of a binary system.

\begin{acknowledgements}
The DASCH project at Harvard is grateful for partial support from NSF grants AST-0407380, AST-0909073, and AST-1313370.
This paper makes use of data from the DR1 of the WASP data \citep{butters2010} as provided by the WASP consortium, and the computing and storage facilities at the CERIT Scientific Cloud, reg. no. CZ.1.05/3.2.00/08.0144, which is operated by Masaryk University, Czech Republic. This research has made use of NASA's Astrophysics Data System. Work on the paper has been supported by LH14300. MS acknowledges the support of the postdoctoral fellowship programme of the Hungarian Academy of Sciences at the Konkoly Observatory as a host institution. We are very grateful to the anonymous referee, who significantly helped to improve this paper.
\end{acknowledgements}


\begin{appendix}\label{appendix1}

\section{Application of the non-linear LSM on calculating the LiTE}\label{liteappendix1}
We assumed a group of $n$ maxima timings given in Heliocentric Julian Date\footnote{Times of maxima in Barycentric Julian Date should be used, but the barycentric correction is lower than the accuracy of maxima times from the GEOS database, which contains times in HJD valid to the fourth decimal place. The accuracy of the determined maxima is likewise mostly lower -- especially for photographic or visual measurements.} (HJD) determined from observations. Times are inserted in the column vector $\bf y$ with a size $n \times 1$. Each time of an observed maximum $T_{l}$ has a corresponding uncertainty $\sigma_{l}$. We assumed that the quality of the $l$-measurement can be quantified by the $l$-weight using the
relation $w_{l}=\sigma_{l}^{-2}$ , and these weights were inserted in the column vector $\bf{w}$. For a correct calculation, the
weights were normalised (the average value of weights is $\overline{w}=1$) and were inserted in a square matrix ${\bf W}={\rm diag}({\bf w})$.

In the next step, we selected the equation for the model function $T_{\rm cal}(T,{\bf b})$ with unknown parameters in vector ${\bf b}$ . Changes in the position times of the maxima for a pulsating star that are caused by the \lt~(periodic changes in the \oc~diagram) can be expressed by
\begin{equation} 
\small
T_{\rm cal}(T,{\bf b})= M_{0} + P_{\rm puls}\times N + \mathnormal{\Delta},
\label{eq:liteEphLinA}
\end{equation}
\noindent where $M_{0}$ is the zero epoch of pulsation in HJD, $P_{\rm puls}$ is the pulsation period in days, $N$ is the number of pulsation cycles from $M_{0}$, and parameter $\mathnormal{\Delta}$ is the correction for the \lt. The integer number of the pulsation cycle (epoch) $N$ is
\begin{equation}
\small
N = \textrm{round}\left(\frac{T - M_{0}}{P_{\rm puls}} \right).
\end{equation}

\noindent An optional more complex model that includes the parabolic trend in the \oc~diagram \citep[e.g.][]{zhu2012} uses the modified Eq.~(\ref{eq:liteEphLinA}) in the form of
\begin{equation}
\small
T_{\rm cal}(T,{\bf b}) = M_{0} + P_{\rm puls}\times N + \frac{1}{2}\,P_{\rm puls}\,\dot{P}_{\rm puls}\times N^2 +\mathnormal{\Delta},
\label{eq:liteEphParA}
\end{equation}
\noindent where $P_{\rm puls}$ is the instantaneous pulsation period at the moment $M_{0}$, and parameter $\dot{P}_{\rm puls} = {\rm d}P_{\rm puls}/{\rm d}t$ is the rate of changes in the pulsation period.

The correction $\mathnormal{\Delta}$ for the \lt~includes calculating the orbit of the pulsating star around the binary mass centre in relative units and is given by the equation adopted from \citep{irwin1952a}
\begin{equation} 
\small
\mathnormal{\Delta} = A \,\left[ (1-e^{2})\,\frac{\sin(\nu+\omega)}{1+e\,\cos\nu} + e\,\sin\omega\right],
\end{equation}
\noindent where $e$ is the numerical eccentricity, $\nu$ the
true anomaly, $\omega$ the argument of periastron in degrees, and $A$ is a constant in light days, which compares the shift in a radial position to the time delay caused by the constant speed of light. The true anomaly $\nu$ is calculated from
\begin{equation} 
\small
\tan\frac{\nu}{2} = \sqrt{ \frac{1+e}{1-e} } \tan\frac{E}{2},
\end{equation}
and the eccentric anomaly $E$ is iteratively determined in our code by Newton's method from Kepler's equation
\begin{equation} 
\small
E = M + e\,\sin E.
\end{equation}
The mean anomaly $M$ is in the form
\begin{equation}
\small
M = \frac{2\,\pi\,(T-T_{0})} {P_{\rm orbit}},\end{equation}
\noindent where the orbital period $P_{\rm orbit}$ is in days
and the time of periastron passage $T_{0}$ in HJD.

The constant $A$ is the projection of the semi-major axis of the pulsating component $a_{1}$ on the unit light day 
\begin{equation}
\small
A = \frac{a_{1}\,\sin i\, {\rm au}}{86400\,c}\doteq\frac{a_{1}\,\sin i}{173.145},
\end{equation}
\noindent where $i$ is the inclination angle of the orbit in degrees, au is length of the astronomical unit in metres, $c$ is the speed of the light in vacuum in m\,s$^{-1}$. The semi-amplitude of \lt~changes in the \oc~diagram in days is then 
\begin{equation}
\small
A_{\rm LiTE} = A\,\sqrt{1-e^{2}\,\cos^{2}\omega}.
\end{equation}

Subsequently, the observed values of the time of maximum can be compared with those from the model obtained from Eqs.~\ref{eq:liteEphLinA} and \ref{eq:liteEphParA}. Their difference for a given set of parameters is equal to
\begin{equation} 
\small
\delta T_{l} = T_{l} - T_{{\rm cal},l}(T_{l},{\bf b}).
\label{eq:differ}
\end{equation}
\noindent The LSM described in detail in \citep{mikulasek2013} points that the best model has the lowest sum of squares of residuals between observation and model. The
modified form of the LSM used the weighted form
\begin{equation} 
\small
\delta T_{{\rm mod}, l} = \frac{\delta T_{l}}{\sigma_{l}}= \frac{T_{l} - T_{{\rm cal},l}(T_{l},{\bf b})}{\sigma_{l}},
\label{eq:differWeight}
\end{equation} 
\noindent and then the sum is 
{\small{\setlength\arraycolsep{2pt}
\begin{eqnarray}
\chi^{2}(\textbf{b}) = \sum_{l = 1}^{n} {\delta T}^{2}_{{\rm mod}, l} =  \sum_{l = 1}^{n} \left[ \frac{T_{l} - T_{{\rm cal},l}(T_{l},{\bf b})}{\sigma_{l}}  \right]^{2} =
\nonumber\\
=\sum_{l = 1}^{n} \left[ \delta T_{l}^{2}\,w_{l} \right].
\end{eqnarray}}}

\noindent Its normalised value $\chi_{\rm R}^{2}(\textbf{b}_{k}) = \chi^{2}(\textbf{b}_{k})/(n-g)$, where $n$ is the number of measurements and $g$ is the number of free parameters (the length of matrix $\textbf{b}$), was used as an indicator of the quality of the $k$-fit.

The first step of the non-linear LSM is linearisation of the non-linear model function (Eqs.~\ref{eq:liteEphLinA} or \ref{eq:liteEphParA}) by a Taylor decomposition of the first order \citep[see][]{mikulasek2006,mikulasek2011a}
\begin{equation}
\small
T_{\rm cal} \cong T_{\rm cal}(T,\textbf{b}_{0}) + \sum_{j=1}^{g} \Delta b_{j}\frac{\partial T_{\rm cal}(T,\textbf{b}_{0})}{\partial b_{j}},
\end{equation}
\noindent where $b_{j}$ are individual free parameters from the vector $\textbf{b}$, vector $\textbf{b}_{0}$ contains the initial estimates of the parameters, and $\Delta b_{j}$ are their corrections in the vector ${\boldsymbol \Delta} {\bf b}$. 

Important equations for solving non-linear LSM by matrices are
{\small
\begin{equation}
\begin{aligned}
{\bf U} = {\bf X^{T}\,W\,{\boldsymbol \Delta} y},\quad {\bf V} = {\bf X^{T}\,W\,X}, \\
{\bf H} = {\bf V}^{-1} = ({\bf X^{T}\,W\,X})^{-1}, \quad {\boldsymbol \Delta} {\bf b} = {\bf H\,U},  
\label{eq:lsm}
\end{aligned}
\end{equation}}

\noindent from which the new parameters in vector $\textbf{b}_{1}$ can be calculated as $\textbf{b}_{1} = \textbf{b}_{0} + {\boldsymbol \Delta} {\bf b}$. The difference between observed values and model is ${\boldsymbol \Delta} {\bf y} = {\bf y} - T_{\rm cal}$. The matrix with the derivatives has the form
{\small{\setlength\arraycolsep{2pt}
\begin{eqnarray}
{\bf X} = \left[\frac{\partial T_{\rm cal}}{\partial M_{0}}, 
\frac{\partial T_{\rm cal}}{\partial P_{\rm puls}}, 
\frac{\partial T_{\rm cal}}{\partial T_{0}}, 
\frac{\partial T_{\rm cal}}{\partial P_{\rm orbit}},
\frac{\partial T_{\rm cal}}{\partial A},\right. 
\nonumber\\
\left.,\frac{\partial T_{\rm cal}}{\partial \omega}, 
\frac{\partial T_{\rm cal}}{\partial e} \right],
\end{eqnarray}}}

\noindent where individual derivatives are also presented
{\small\begin{equation*}
\frac{\partial T_{\rm cal}}{\partial M_{0}} = 1,\quad
\frac{\partial T_{\rm cal}}{\partial T_{0}} = \frac{\partial\mathnormal{\Delta}}{\partial\nu}\, \frac{\partial\nu}{\partial E}\, \frac{\partial E}{\partial M}\, \frac{\partial M}{\partial T_{0}}, 
\end{equation*}
\begin{equation*} 
\frac{\partial T_{\rm cal}}{\partial P_{\rm puls}} = N, \quad
\frac{\partial T_{\rm cal}}{\partial P_{\rm orbit}} = \frac{\partial\mathnormal{\Delta}}{\partial\nu}\, \frac{\partial\nu}{\partial E}\, \frac{\partial E}{\partial M}\, \frac{\partial M}{\partial P_{\rm orbit}}, 
\end{equation*}
\begin{equation*} 
\frac{\partial T_{\rm cal}}{\partial A} =  \frac{(1-e^{2})\,\sin(\nu+\omega)}{1+e\,\cos\nu} + e\,\sin\omega, 
\end{equation*}
\begin{equation*} 
\frac{\partial T_{\rm cal}}{\partial \omega} =  A \,\left[ \frac{(1-e^{2})\,\cos(\nu+\omega)}{1+e\,\cos\nu} + e\,\cos\omega\right], 
\end{equation*}
\begin{equation*} 
\frac{\partial T_{\rm cal}}{\partial e} = \frac{\partial\mathnormal{\Delta}}{\partial e} + \frac{\partial\mathnormal{\Delta}}{\partial\nu}\, \frac{\partial\nu}{\partial e} + \frac{\partial\mathnormal{\Delta}}{\partial\nu}\, \frac{\partial\nu}{\partial E}\, \frac{\partial E}{\partial e}.
\end{equation*}}

\noindent Other necessary derivatives are

{\small\begin{equation*} 
\frac{\partial\mathnormal{\Delta}}{\partial\nu} = \frac{A\,(1-e^{2}) }{1+e\,\cos\nu}\, \left[ \cos(\nu+\omega) + \frac{ e\,\sin(\nu+\omega)\,\sin\nu}{1+e\,\cos\nu} \right],
\end{equation*}}
{\small\begin{equation*} 
\frac{\partial\mathnormal{\Delta}}{\partial e} = A\, \left\{ \sin\omega - \frac{ \sin(\nu+\omega)\,[2\,e + (1 + e^2)\,\cos\nu] }{ (1+e\,\cos\nu)^2} \right\},
\end{equation*}}
{\small\begin{equation*} 
\frac{\partial\nu}{\partial E} = \sqrt{ \frac{1+e}{1-e} }\, \left( \frac{\cos  \displaystyle\frac{\nu}{2} }{ \cos  \displaystyle \frac{E}{2} } \right)^{\!\!2},\quad
\frac{\partial\nu}{\partial e} = \frac{\sin\nu}{ 1-e^{2} } ,
\end{equation*}}
{\small\begin{equation*} 
\frac{\partial E}{\partial M} = \frac{1}{1-e\,\cos E},\quad
\frac{\partial E}{\partial e} = \frac{\sin E}{ 1-e\,\cos E},
\end{equation*}}
{\small\begin{equation*} 
\frac{\partial M}{\partial P_{\rm orbit}} = \frac{-2\,\pi\,(T-T_{0})}{P_{\rm orbit}^2}, \quad
\frac{\partial M}{\partial T_{0}} = \frac{-2\,\pi}{P_{\rm orbit}}. 
\end{equation*}}

\noindent Matrix $\textbf{X}$ will be expanded by about one additional member to calculate the parabolic trend according to Eq.~\ref{eq:liteEphParA}
{\small{\setlength\arraycolsep{2pt}
\begin{eqnarray}
{\bf X} = \left[\frac{\partial T_{\rm cal}}{\partial M_{0}}, 
\frac{\partial T_{\rm cal}}{\partial P_{\rm puls}}, 
\frac{\partial T_{\rm cal}}{\partial T_{0}}, 
\frac{\partial T_{\rm cal}}{\partial P_{\rm orbit}},
\frac{\partial T_{\rm cal}}{\partial A},\right. 
\nonumber\\
\left.,\frac{\partial T_{\rm cal}}{\partial \omega}, 
\frac{\partial T_{\rm cal}}{\partial e}, \frac{\partial T_{\rm cal}}{\partial \dot{P}_{\rm puls}} \right],
\end{eqnarray}}}

\noindent where two of $\textbf{X}$ members are in the form
\begin{equation*}
\small
\frac{\partial T_{\rm cal}}{\partial P_{\rm puls}} = N + \frac{1}{2}\,\dot{P}_{\rm puls}\times N^2 ,\quad
\frac{\partial T_{\rm cal}}{\partial \displaystyle\dot{P}_{\rm puls}} = \frac{1}{2}\, P_{\rm puls} \times N^2. 
\end{equation*}

The determined parameters allow calculating the radial velocity (RV) changes caused by the secondary component \citep[e.g.][]{irwin1952b}
\begin{equation} 
\small
RV_{1} = \gamma + K_{1}\left[\cos(\nu + \omega) + e\,\cos\omega\right],
\end{equation}
\noindent where $\gamma$ is systematic velocity mass-centre of the binary system from the Sun in km\,s$^{-1}$ ($\gamma$-velocity) and $K_{1}$ is the semi-amplitude of RV changes in km\,s$^{-1}$ given by 
\begin{equation}
\small
K_{1} = \frac{2\,\pi\,a_{1}\,\sin i\,{\rm au}}{8.64\times10^{7}\,P_{\rm orbit}\sqrt{(1-e^{2})}},
\end{equation}
\noindent where the projection of the semi-major axis $a_{1}\,\sin i$ is in au, the constant au in meters, and the orbital period $P_{\rm orbit}$ in days.

\end{appendix}


\begin{thebibliography}{}
        \bibitem[Abt(1970)]{abt1970} Abt, H.~A.\ 1970, \apjs, 19, 387

        \bibitem[Arellano Ferro et al.(2013)]{arellanoferro2013} Arellano Ferro, A., Pe{\~n}a, J.~H., \& Figuera Jaimes, R.\ 2013, \rmxaa, 49, 53

        \bibitem[Barnes et al.(1988)]{barnes1988} Barnes, T.~G.~III., Frueh, M.~L., Moffett, T.~J. et al.\ 1988, \apjs, 67, 403

        \bibitem[Boenigk(1958)]{boenigk1958} Boenigk, T.\ 1958, \actaa, 8, 13 

        \bibitem[Boninsegna et al.(2002)]{boninsegna2002} Boninsegna, R., Vandenbroere, J., Le Borgne, J.~F., \& Geos Team 2002, IAU Colloq.~185: Radial and Nonradial Pulsations as Probes of Stellar Physics, 259, 166 

        \bibitem[Burd et al.(2004)]{burd2004} Burd, A., Cwiok, M., Czyrkowski, H., et al.\ 2004, Astronomische Nachrichten, 325, 674

        \bibitem[Butters et al.(2010)]{butters2010} Butters O.~W., West, R.~G., Anderson, D.~R. et al. 2010 \aap, 520, 10

        \bibitem[Chandler(1888)]{chandler1888} Chandler, S.~C. \ 1888 \aj, 7, 165

        \bibitem[Chrastina(2013)]{chrastina2013} Chrastina, M. \ 2013, Study of fast light changes of interacting binaries, Ph.D. Thesis, DTPA of Masaryk University, Brno, Czech Republic

        \bibitem[ESA(1997)]{esa1997} ESA 1997, The Hipparcos and Tycho Catalogs, SP--1200

        \bibitem[Fernley(1993)]{fernley1993} Fernley, J.\ 1993, \aap, 268, 591

        \bibitem[Fernley \& Barnes(1997)]{fernley1997} Fernley, J., \& Barnes, T.~G.\ 1997, \aaps, 125, 313 

        \bibitem[Grindlay et al.(2009)]{grindlay2009} Grindlay, J., Tang, S., Simcoe, R. et al. \ 2009, ASPC, 410, 101

        \bibitem[Guthnick \& Prager(1929)]{guthnick1929} Guthnick, P., \& Prager, R.,\ 1929, Beob. Zirk., 11, 32

        \bibitem[Hajdu et al.(2015)]{hajdu2015} Hajdu, G., Catelan, M., Jurcsik, J., et al.\ 2015, \mnras, 449, L113

        \bibitem[Hajdu(2015, priv. comm.)]{hajdu2015b} Hajdu, G., \ 2015, priv. comm. 

        \bibitem[Hemenway(1975)]{hemenway1975} Hemenway, M.~K.\ 1975, \aj, 80, 199 

        \bibitem[H\"{u}bscher(2014)]{hubscher2014} H\"{u}bscher, J.\ 2014, Information Bulletin on Variable Stars, 6118, 1 
                                        
        \bibitem[H\"{u}bscher \& Lehmann(2015)]{hubscher2015} H\"{u}bscher, J., \& Lehmann, P.~B.\ 2015, Information Bulletin on Variable Stars, 6149, 1 

        \bibitem[Irwin(1952a)]{irwin1952a} Irwin, J.~B.\ 1952a, \apj, 116, 211

        \bibitem[Irwin(1952b)]{irwin1952b} Irwin, J.~B.\ 1952b, \apj, 116, 218

        \bibitem[Kiss et al.(1995)]{kiss1995} Kiss, L.~L., Szatmary, K., Gal, J., \& Kaszas, G.\ 1995, Information Bulletin on Variable Stars, 4205, 1 

        \bibitem[Layden(1993)]{layden1993} Layden, A.~C.\ 1993, The Metallicities and Kinematics of the Local RR Lyrae Variables, Ph.D.~Thesis, Yale University, New Haven, USA 

        \bibitem[Layden(1994)]{layden1994} Layden, A.~C.\ 1994, \aj, 108, 1016 

        \bibitem[Le Borgne et al.(2007)]{leborgne2007} Le Borgne, J.~F., Paschke, A., Vandenbroere, J., et al.\ 2007, \aap, 476, 307

        \bibitem[Lee et al.(2010)]{lee2010} Lee, J.~W., Kim, Ch.-H., Kim, D.~H., et al.\ 2010, \aj, 139, 2669

        \bibitem[Li \& Qian(2014)]{li2014} Li, L.-J., Qian, S.-B.,\ 2014, \mnras, 444, 600

        \bibitem[Liakos \& Niarchos(2011a)]{liakos2011a} Liakos, A., \& Niarchos, P.,\ 2011a, Information Bulletin on Variable Stars, 6099, 1 

		\bibitem[Liakos \& Niarchos(2011b)]{liakos2011b} Liakos, A., \& Niarchos, P.,\ 2011b, Information Bulletin on Variable Stars, 5990, 1 

        \bibitem[Li\v{s}ka \& Li\v{s}kov\'{a}(2014)]{liska2014a} Li\v{s}ka, J., Li\v{s}kov\'{a}, Z.,\ 2014, Information Bulletin on Variable Stars, 6124, 1 

        \bibitem[Li\v{s}ka (2014)]{liska2014b} Li\v{s}ka, J.,\ 2014, Information Bulletin on Variable Stars, 6119, 1 

        \bibitem[Li\v{s}ka et al.(2015)]{liska2015} Li\v{s}ka, J., Skarka, M., Zejda, M., \& Mikul\'{a}\v{s}ek, Z.\ 2015, arXiv:1504.05246

        \bibitem[Liu \& Janes(1989)]{liu1989} Liu, T., \& Janes, K.~A.\ 1989, \apjs, 69, 593

        \bibitem[Liu \& Janes(1990)]{liu1990} Liu, T., \& Janes, K.~A.\ 1990, \apj, 354, 273
        \bibitem[Maintz(2005)]{maintz2005} Maintz, G.\ 2005, \aap, 442, 381

        \bibitem[Mikul{\'a}{\v s}ek et al.(2006)]{mikulasek2006} Mikul{\'a}{\v s}ek, Z., Wolf, M., Zejda, M., \& Pecharov{\'a}, P.\ 2006, \apss, 304, 363 

        \bibitem[Mikul\'{a}\v{s}ek \& Gr\'{a}f(2011)]{mikulasek2011a} Mikul\'{a}\v{s}ek, Z. \& Gr\'{a}f, T.\ 2011, arXiv:1111.1142

        \bibitem[Mikul\'{a}\v{s}ek et al.(2011)]{mikulasek2011b} Mikul{\'a}{\v s}ek, Z., {\v Z}i{\v z}{\v n}ovsk{\'y}, J., Zejda, M., et al.\ 2011, in Magnetic Stars, Proceedings of the International Conference, held in the Special Astrophysical Observatory of the Russian AS, August 27- September 1, 2010, Eds: I. I. Romanyuk and D. O. Kudryavtsev, 431

        \bibitem[Mikul{\'a}{\v s}ek \& Zejda(2013)]{mikulasek2013} Mikul{\'a}{\v s}ek, Z., \& Zejda, M.\ 2013, in \'{U}vod do studia prom\v{e}nn\'{y}ch hv\v{e}zd, ISBN 978-80-210-6241-2, Masaryk University, Brno

        \bibitem[Oke et al.(1962)]{oke1962} Oke, J.~B., Giver, L.~P., Searle, L.\ 1962, \apj, 136, 393

        \bibitem[Paschke \& Br\'{a}t(2006)]{paschke2006} Paschke, A., Br\'{a}t, L.\ 2006, OEJV, 23, 13

        \bibitem[Panchatsaram(1981)]{panchatsaram1981} Panchatsaram, T.\ 1981, \apss, 77, 179

        \bibitem[Payne-Gaposchkin(1939)]{payne-gaposchkin1939} Payne-Gaposchkin, C.,\ 1939, Proceedings of the American Philosophical Society, 81, 189

        \bibitem[Payne-Gaposchkin(1954)]{payne-gaposchkin1954} Payne-Gaposchkin, C.\ 1954, Annals of Harvard College Observatory, 113, 151 

        \bibitem[Pietrzy{\'n}ski et al.(2012)]{pietrzynski2012} Pietrzy{\'n}ski, G., Thompson, I.~B., Gieren, W., et al.\ 2012, \nat, 484, 75 

        \bibitem[Pollacco et al.(2006)]{pollacco2006} Pollacco, D.~L. , Skillen, I., Collier Cameron, A., et al. \ 2006, PASP, 118, 1407

        \bibitem[Preston et al.(1961)]{preston1961} Preston, G.~W., Spinrad, H. \& Varsavsky, C.~M.\ 1961, \apj, 133, 484

        \bibitem[Preston \& Paczynski(1964)]{preston1964} Preston, G.~W. \& Paczynski, B.\ 1964, \apj, 140, 181

        \bibitem[Pribulla et al.(2000)]{pribulla2000} Pribulla, T., Chochol, D., Tremko, J. et al. \ 2000, Contributions of the Astronomical Observatory Skalnate Pleso, 30, 117

        \bibitem[Pr{\v s}a et al.(2008)]{prsa2008} Pr{\v s}a, A., Guinan, E.~F., Devinney, E.~J., \& Engle, S.~G.\ 2008, \aap, 489, 1209

        \bibitem[Robinson \& Shapley(1940)]{robinson1940} Robinson, L.~V., \& Shapley, H.\ 1940, Annals of Harvard College Observatory, 90, 27 
        
        \bibitem[Saha \& White(1990a)]{saha1990a} Saha, A., \& White, R.~E.\ 1990a, \pasp, 102, 148    

        \bibitem[Saha \& White(1990b)]{saha1990b} Saha, A., \& White, R.~E.\ 1990b, \pasp, 102, 495, Erratum 
        
        \bibitem[Sanford(1949)]{sanford1949} Sanford, R.~F.\ 1949, \apj, 109, 208    

        \bibitem[Sesar(2012)]{sesar2012} Sesar, B.\ 2012, \aj, 144, 114
                                        
        \bibitem[Siudek et al.(2011)]{siudek2011} Siudek, M., Malek, K., Mankiewicz, L., et al.\ 2011, Acta Polytechnica, 51, 68
        
        \bibitem[Skarka(2014)]{skarka2014} Skarka, M.\ 2014, \mnras, 445, 1584

        \bibitem[Slawson et al.(2011)]{slawson2011} Slawson, R.~W., Prsa, A., Welsh, W. F., et al.\ 2011, \aj, 142, 160
        
        \bibitem[Smolec et al.(2013)]{smolec2013} Smolec, R., Pietrzy{\'n}ski, G., Graczyk, D., et al.\ 2013, \mnras, 428, 3034

        \bibitem[Solano et al.(1997)]{solano1997} Solano, E., Garrido, R., Fernley, J., \& Barnes, T.~G.\ 1997, \aaps, 125, 321
        
        \bibitem[Soszy{\'n}ski et al.(2003)]{soszynski2003} Soszy{\'n}ski, I., Udalski, A., Szymanski, M., et al.\ 2003, \actaa, 53, 93
        
        \bibitem[Soszy{\'n}ski et al.(2009)]{soszynski2009} Soszy{\'n}ski, I., Udalski, A., Szyma{\'n}ski, M.~K., et al.\ 2009, \actaa, 59, 1
                                        
        \bibitem[Soszy{\'n}ski et al.(2011)]{soszynski2011} Soszy{\'n}ski, I., Dziembowski, W.~A., Udalski, A., et al.\ 2011, \actaa, 61, 1 
                
        \bibitem[Szatm\'{a}ry(1990)]{szatmary1990} Szatm\'{a}ry, K.\ 1990, Journal of the American Association of Variable Star Observers, 19, 52

        \bibitem[Szeidl et al.(1986)]{szeidl1986} Szeidl, B., Olah, K., \& Mizser, A.\ 1986, Communications of the Konkoly Observatory Hungary, 89, 57

        \bibitem[Van Hamme \& Wilson(2007)]{vanhamme2007} Van Hamme, W., \& Wilson, R.~E.\ 2007, \apj, 661, 1129

        \bibitem[Wade et al.(1992)]{wade1992} Wade, R.~A., Saha, A., White, R.~E., \& Fried, R.\ 1992, Bulletin of the American Astronomical Society, 24, 1225

        \bibitem[Wade et al.(1999)]{wade1999} Wade, R.~A., Donley, J., Fried, R., et al.\ 1999, \aj, 118, 2442 

        \bibitem[Watson et al.(2006)]{watson2006} Watson, C.~L., Henden, A.~A. \& Price, A.\ 2006, Society for Astronomical Sciences Annual Symposium, 25, 47  

        \bibitem[Wilson \& Van Hamme(2014)]{wilson2014} Wilson, R.~E., \& Van Hamme, W.\ 2014, \apj, 780, 151 

        \bibitem[Wolf et al.(1999)]{wolf1999} Wolf, M., \v{S}arounov\'{a}, L., Bro\v{z}, M., Horan, R.,\ 1999, Information Bulletin on Variable Stars, 4683, 1

        \bibitem[Wolf et al.(2007)]{wolf2007} Wolf, M., Kotkov\'{a}, L., Br\'{a}t, L., et al.\ 2007, Information Bulletin on Variable Stars, 5780, 1

        \bibitem[Woltjer(1922)]{woltjer1922} {Woltjer}, Jr., J. \ 1922, \bain, 1, 93
                                        
        \bibitem[Wo{\'z}niak et al.(2004)]{wozniak2004} Wo{\'z}niak, P.~R., Vestrand, W.~T., Akerlof, C.~W., et al. \ 2004, \aj, 127, 2436

        \bibitem[Zasche(2008)]{zasche2008} Zasche, P. \ 2008, arXiv:0801.4258

        \bibitem[Zhou(2010)]{zhou2010} Zhou, A.~-Y.\ 2010, arXiv:1002.2729v6 

        \bibitem[Zhu et al.(2012)]{zhu2012} Zhu, L.-Y., Zejda, M., Mikul{\'a}{\v s}ek, Z., et al.\ 2012, \aj, 144, 37 
                                        
\end{thebibliography}
\end{document}